\documentclass[11pt]{article}
\usepackage{jcapmod}

\usepackage[normalem]{ulem}

\usepackage{booktabs}
\usepackage[english]{babel}
\usepackage{amsmath,amssymb,amsbsy,amstext, amsthm, simplewick}
\usepackage{hyperref}
\usepackage{graphicx}
\usepackage{amsfonts}
\usepackage{amssymb}
\usepackage[small]{caption}
\usepackage{upgreek}
  \usepackage{framed}
\usepackage{enumitem}
  \usepackage{fontawesome}
  \usepackage{jcapmod}
\usepackage{booktabs}
\usepackage[english]{babel}
\usepackage{amsmath,amssymb,amsbsy,amstext, amsthm, simplewick}
\usepackage{wrapfig}
\usepackage{hyperref}
\usepackage{graphicx}
\usepackage{amsfonts}
\usepackage{amssymb}
\usepackage{subfig}
\usepackage{upgreek}
 \usepackage{exscale,relsize}
 \usepackage[makeroom]{cancel}
\usepackage{soul}
\usepackage{bbold}
\usepackage{lipsum}
\usepackage{mdframed}
\usepackage{mathtools}
\usepackage[dvipsnames]{xcolor}
\usepackage{tikz}
\usepackage{tikz-cd}
\usepackage[export]{adjustbox}

 \newcommand{\circled}[2][]{%
  \tikz[baseline=(char.base)]{%
    \node[shape = circle, draw, inner sep = 1pt]
    (char) {\phantom{\ifblank{#1}{#2}{#1}}};%
    \node at (char.center) {\makebox[0pt][c]{#2}};}}
\robustify{\circled}

\newcommand*\diff{\mathop{}\!\mathrm{d}}
\newcommand{\overbar}[1]{\mkern 1.5mu\overline{\mkern-1.5mu#1\mkern-1.5mu}\mkern 1.5mu}

\usepackage{colortbl}
\definecolor{lightgreen}{cmyk}{0.2, 0, 0.2, 0.2}
\definecolor{lightgray}{cmyk}{0.1,0.2,0,0.1}
\definecolor{lightgray2}{cmyk}{0.1,0.1,0,0.1}

\makeatletter
\newlength{\apb@width}
\newcommand{\autoparbox}[2][c]{\settowidth{\apb@width}{#2}\parbox[#1]{\apb@width}{#2}}

\newcommand{\Cen}[2]{%
  \ifmeasuring@
    #2%
  \else
    \makebox[\ifcase\expandafter #1\maxcolumn@widths\fi]{$\displaystyle#2$}%
  \fi
}
\makeatother

\newcommand{\beq}{\begin{equation}\begin{aligned}}
\newcommand{\eeq}{\end{aligned}\end{equation}}

\newcommand{\CL}{\texttt{$\mathcal{C}\text{osmo}\mathcal{L}\text{attice}$}}

\numberwithin{equation}{section}

\def\beq{\begin{equation}}
\def\eeq{\end{equation}}

\def\Beq{\begin{equation}\begin{aligned}}
\def\Eeq{\end{aligned}\end{equation}}

\def\bea{\begin{eqnarray}}
\def\eea{\end{eqnarray}}

\def\beq{\begin{equation}}
\def\eeq{\end{equation}}
\def\bea{\begin{eqnarray}}
\def\eea{\end{eqnarray}}

\def\bp{\boldsymbol{p}}

\def\bx{\boldsymbol{x}}
\def\bq{\boldsymbol{q}}

\DeclareRobustCommand{\SkipTocEntry}[4]{}
\DeclareSymbolFont{extraup}{U}{zavm}{m}{n}
\DeclareMathSymbol{\varheart}{\mathalpha}{extraup}{86}
\DeclareMathSymbol{\vardiamond}{\mathalpha}{extraup}{87}

\begin{document}

\hypersetup{pageanchor=false}

\begin{titlepage}

\setcounter{page}{1} \baselineskip=15.5pt \thispagestyle{empty}

\bigskip\

\begin{flushright}
\mbox{} 
DESY-24-058 
\end{flushright}

\vspace{0.3cm}
\begin{center}

{\fontsize{20.74}{24}\selectfont  \sffamily \bfseries Gravitational wave signatures of \\

\vspace{0.2cm} post-fragmentation reheating}

\end{center}

\vspace{0.2cm}

\begin{center}
{\fontsize{12}{30}\selectfont  Marcos A.~G.~Garcia$^{\spadesuit}$\footnote{marcos.garcia@fisica.unam.mx} and Mathias Pierre$^{\clubsuit}$\footnote{mathias.pierre@desy.de}}
\setcounter{footnote}{0}
\end{center}

\begin{center}

\vskip 7pt

\textsl{$^{\spadesuit}$ Departamento de F\'isica Te\'orica, Instituto de F\'isica, Universidad Nacional Aut\'onoma de M\'exico, Ciudad de M\'exico C.P. 04510, Mexico}\\
\textsl{$^{\clubsuit}$ Deutsches Elektronen-Synchrotron DESY, Notkestr. 85, 22607 Hamburg, Germany}
\vskip 7pt

\end{center}

\vspace{0.3cm}
\centerline{\bf ABSTRACT}
\vspace{0.3cm}

\noindent
After cosmic inflation, coherent oscillations of the inflaton field about a monomial potential $V(\phi)\sim \phi^k$ result in an expansion phase characterized by a stiff equation-of-state $w\simeq(k-2)/(k+2)$. Sourced by the oscillating inflaton condensate, parametric (self)resonant effects can induce the exponential growth of inhomogeneities eventually backreacting and leading to the fragmentation of the condensate. In this work, we investigate realizations of inflation giving rise to such dynamics, assuming an inflaton weakly coupled to its decay products. As a result, the transition to a radiation-dominated universe, i.e. reheating, occurs after fragmentation. We estimate the consequences on the production of gravitational waves by computing the contribution induced by the stiff equation-of-state era in addition to the signal generated by the fragmentation process for $k=4,6,8,10$. We find that the signal generated during the fragmentation process gives a larger contribution than the one induced by the stiff equation-of-state era in given frequency ranges for all values of $k$. Our results are independent of the reheating temperature provided that reheating is achieved posterior to fragmentation. Our work shows that the dynamics of such weakly-coupled inflaton scenario can actually result in characteristic gravitational wave spectra with frequencies from Hz to GHz, in the reach of future gravitational wave observatories, in addition to the complementarity between upcoming detectors in discriminating (post)inflation scenarios. We advocate the need of developing high-frequency gravitational wave detectors to gain insight into the dynamics of inflation and reheating.

\vspace{0.3cm}
\noindent 
\vfill

\begin{flushleft}
{September} 2024
\end{flushleft}

 \end{titlepage}

\hypersetup{pageanchor=true}

\tableofcontents

\section{Introduction}

The inflationary paradigm, which predicts the existence of an early epoch of accelerated expansion of the universe, provides one of the most compelling solutions to the initial conditions problems of the standard Big Bang cosmological model.\footnote{For reviews, see~\cite{Olive:1989nu,Linde:1990flp,Lyth:1998xn,Linde:2000kn}.} In its arguably simplest realization, the quasi-exponential growth of the universe is driven by the slow roll of a spatially homogeneous scalar field (the inflaton) over a sufficiently flat potential. The excitation of the quantum fluctuations of this field and the spacetime metric would source an almost scale-invariant power spectrum of adiabatic scalar fluctuations, which is in remarkable agreement with Cosmic Microwave Background (CMB) and Large Scale Structure observations. Nevertheless, what is considered the smoking gun signal for inflation, a primordial spectrum of parity odd ($B$-mode) polarization fluctuations in the CMB, sourced by tensor modes excited during inflation, has so far escaped the capabilities of existing CMB observatories. Currently, the bound on the ratio of the tensor and scalar spectra at the pivot scale $k_*=0.05\,{\rm Mpc}^{-1}$ is $r< 0.036$, at the 95\% CL~\cite{BICEP:2021xfz}. The non-observation of such primordial tensor modes seems to favor models with concave, asymptotically flat potentials~\cite{Planck:2018jri}, among which the Starobinsky~\cite{Starobinsky:1980te}, Higgs~\cite{Bezrukov:2007ep}, and $\alpha$-attractor models~\cite{Ellis:2013nxa,Kallosh:2013yoa,Kallosh:2013maa,Kallosh:2013hoa,Ellis:2019bmm} are theoretically well motivated examples.

The superhorizon stretching of vacuum tensor modes during the slow roll of the inflaton is however not the only source of gravitational waves predicted by inflation. The epoch of accelerated expansion must eventually come to an end, after which the energy density of the inflaton is to be transferred to the Standard Model (and possibly dark matter) degrees of freedom, during what is known as the reheating epoch. In the slow roll picture, this ``graceful exit'' mechanism is built in, where the coherent oscillations of the inflaton field about the minimum of its potential source particle production via the coupling of the inflaton to light degrees of freedom~\cite{Abbott:1982hn,Shtanov:1994ce,Kofman:1994rk,Ichikawa:2008ne,Amin:2014eta,Kainulainen:2016vzv,Garcia:2020wiy,Garcia:2020eof}. Although for small couplings this dissipation process can be modeled perturbatively, for large couplings collective bosonic enhancement (or fermionic blocking) effects require the use of non-perturbative techniques. For bosonic fields, this {\em preheating} leads to the exponential growth of the fluctuations of the decay products, due to the non-adiabatic change of their effective masses, manifested as parametric resonance~\cite{Shtanov:1994ce,Dolgov:1989us,Traschen:1990sw,Boyanovsky:1995ud,Yoshimura:1995gc,Kofman:1997yn}. If this growth is sustained, the coherent inflaton condensate will eventually be fragmented, due to the large gradients sourced by mode-mode couplings (backreaction)~\cite{Amin:2014eta,Garcia-Bellido:2002fsq,Felder:2006cc,Frolov:2010sz,Garcia:2021iag,Figueroa:2016wxr}. Importantly, these large amplitude gradients can in turn efficiently source a spectrum of high frequency gravitational waves~\cite{Easther:2006gt,Dufaux:2007pt,Garcia-Bellido:2007nns,Garcia-Bellido:2007fiu,Dufaux:2008dn,Dufaux:2010cf,Bethke:2013aba,Bethke:2013vca,Figueroa:2017vfa,Adshead:2018doq,Adshead:2019lbr,Cosme:2022htl,Ringwald:2022xif}. On top of this efficient mechanism, gravitational waves can also be sourced directly by the oscillating condensate~\cite{Choi:2024ilx}, or from graviton bremsstrahlung during reheating, from inflaton decay or inflaton annihilation~\cite{Nakayama:2018ptw,Huang:2019lgd,Barman:2023ymn,Bernal:2023wus,Barman:2023rpg}. Additionally, if certain conditions are met during the preheating epoch, long-lived solitonic-like configurations of the inflaton field called oscillons can be formed through inflaton fragmentation during reheating and can also yield a significant gravitational wave signal~\cite{Antusch:2016con,Lozanov:2019ylm,Lozanov:2022yoy}. Hence, the primordial stochastic gravitational wave background is a potential powerful probe of post-inflationary dynamics~\cite{Ringwald:2022xif}.

In this work we revisit the production of gravitational waves due to the resonant growth of inflaton fluctuations due to its self-resonance. Namely we consider models in which the potential of the inflaton $\phi$ can be approximated as $V(\phi)\propto \phi^k$ (with $k$ even) near its minimum. For $k>2$, the self-interaction of the inflaton will eventually fragment the homogeneous inflaton condensate after the end of inflation, unless reheating can be completed promptly. After fragmentation, the universe is dominated by the relativistic free inflatons, which must subsequently decay into the Standard Model primordial plasma. This implies that the equation of state of the universe transitions as $w\rightarrow -1/3 \rightarrow \frac{k-2}{k+2} \rightarrow 1/3$ from the end of inflation to the end of backreaction. Here we will work under this late reheating assumption, taking into account the evolution of the equation of state parameter and the post-fragmentation completion of reheating, recently studied in~\cite{Garcia:2023eol,Garcia:2023dyf,Garcia:2024rwg}, to determine the present-day amplitude of the stochastic gravitational wave background sourced by the inflaton self-resonance. Moreover, although the super-horizon fluctuations sourced during inflation are generically impervious to the nonlinear preheating dynamics~\cite{Jedamzik:1999um,Ivanov:1999hz,Liddle:1999hq}, modes that re-enter the horizon prior to the end of backreaction will experience a relative enhancement due to the stiffer-than-radiation equation of state parameter~\cite{Giovannini:1998bp,Sahni:2001qp,Tashiro:2003qp,Boyle:2007zx,Caprini:2018mtu,Figueroa:2018twl,Figueroa:2019paj}, which we also quantify here. By means of numerical computation and lattice techniques, we keep track of the time-dependence of the equation-of-state parameter from a stage deep into the inflation era until the completion of the fragmentation process. We use both analytical and lattice techniques to estimate the total resulting gravitational wave signal that we confront to prospects from potential future gravitational wave observatories.

This paper is organized as follows: In Section~\ref{sec:sec2} we review the transition from the inflation slow roll phase (Sec.~\ref{sec:sec21}) to the coherent oscillation of the homogeneous field (Sec.~\ref{sec:sec22}), the growth of inhomogeneities sourced by parametric resonances (Sec.~\ref{sec:sec23}), the numerical simulation of backreaction (Secs.~\ref{sec:sec24} and \ref{sec:sec25}), until the eventual completion of reheating via inflaton decays (Sec.~\ref{sec:sec26}). Section~\ref{sec:gws} contains the main results of this work, and is devoted to the computation of the primordial gravitational wave signal. Sec.~\ref{sec:gws1} presents the resulting energy densities and spectra originated from inflaton self-fragmentation, for a selection of inflaton potentials. In Sec.~\ref{sec:gws2} we address the stochastic spectrum generated from vacuum fluctuation growth during inflation, accounting for the effect of the time-dependent equation of state during reheating. Secs.~\ref{sec:gws3} and \ref{sec:summary} present a brief discussion of additional contributions to the tensor spectrum, and the comparison with observational prospects, respectively. Our summary and conclusions are presented in Section~\ref{sec:conclusions}.

\section{Post-fragmentation reheating: from inflation to a radiation-dominated phase}
\label{sec:sec2}

\subsection{Inflation} \label{sec:sec21}

We consider a phase of cosmic inflation driven by a single inflaton field denoted by $\phi$, minimally coupled to gravity, with a corresponding action of the form
\beq
\label{eq:actionphi}
\mathcal{S} \;=\; \int \diff ^4x\,\sqrt{-g} \left[ -\dfrac{M_P^2}{2} R + \frac{1}{2}(\partial_{\mu}\phi)^2 - V(\phi) + \mathcal{L}_{\rm int} \right] \, ,
\eeq
where  $M_P=1/\sqrt{8\pi\,G}\simeq 2.45\times 10^{18}\,{\rm GeV}$ is the reduced Planck mass, $g$ is the metric determinant and $R$ the Ricci scalar. $\mathcal{L}_{\rm int}$ represents interaction terms between the inflaton and additional species, responsible for reheating. For the inflation model we consider the $\alpha$-attractor T-models described by the potential~\cite{Kallosh:2013hoa} 
\beq\label{eq:attractor}
V(\phi) \;=\; \lambda M_P^4 \left[ \sqrt{6} \tanh \left(\frac{\phi}{\sqrt{6}M_P}\right) \right]^k\, ,
\eeq
which are compatible with the current CMB constraints~\cite{BICEP:2021xfz,Planck:2018jri}, and which appear in the context of no-scale supergravity~\cite{Garcia:2020eof}. The exponent $k$ controls the behavior of the potential close to the minimum
\beq\label{eq:Vphi}
V(\phi) \;\simeq\; \lambda M_P^4 
\left(\frac{\phi}{M_P}\right)^k\,, \qquad \phi \ll M_P\,,
\eeq 
Along with the exponent $k$, the inflaton potential can be fully determined by fixing $\lambda$, which controls the potential at large field values. One can reproduce the central value of the amplitude of the scalar power spectrum, $A_S=e^{3.044}/10^{10}\simeq 2.1\times 10^{-9}$ as measured by {\em Planck}~\cite{Planck:2018vyg}, for the choice of parameter
\begin{equation}\label{eq:lambdapp}
    \lambda \, = \, \frac{18 \pi ^2 A_S}{6^{k/2} N_*^2} \,,
\end{equation}
where $N_*$ is the number of $e$-folds between the horizon-crossing of the CMB fiducial scale $k_*=0.05~\text{Mpc}^{-1}$ (corresponding to $k_*=a_* H_*$), and the end of inflation. It can be related to the field values evaluated at these times, $\phi_*$ and $\phi_{\rm end}$ respectively, as
\begin{equation}
    N_* \;\simeq\; \frac{1}{M_P^2}\int_{\phi_{\rm end}}^{\phi_*} \frac{V(\phi)}{V'(\phi)}\,\diff \phi \;=\; \left( \dfrac{3}{2k} \right) \left[ \cosh \left( \sqrt{\dfrac{2}{3}} \phi_* \right) - \cosh \left( \sqrt{\dfrac{2}{3}} \phi_{\rm end} \right) \right]\,.
    \label{eq:phistar}
\end{equation}
In the following, for definiteness, we chose $N_*=55$.~\footnote{Notice that shifting the value of $N_*$ by up to $10$ $e$-folds would only result in a $\mathcal{O}(1)$ modification in the normalization of $\lambda$ and does not significantly impact our results.} \par \medskip

\noindent
Accelerated expansion is maintained until the end of inflation  when $\ddot{a}=0$, where dots represent differentiation with respect to cosmic time and $a$ is the scale factor, or equivalently when the derivative of the field is $\dot \phi_\text{end}=-\sqrt{V(\phi_\text{end})}$. The field value $\phi_{\rm end}$ can be approximated by~\cite{Ellis:2015pla,Garcia:2020wiy}
\begin{align}
   \phi_{\rm end} &\simeq 
   \sqrt{\frac{3}{8}}M_P\ln\left[\frac{1}{2} + \frac{k}{3}\left(k + \sqrt{k^2+3}\right)\right].
   \label{phiend}
\end{align}
This corresponds to an energy density at the end of inflation
\begin{equation}
    \rho_{\rm end} \, \equiv \, 3 H_\text{end}^2 M_P^2  \sim (4.8\times 10^{15}\,{\rm GeV})^4\,,
    \label{eq:rhoend}
\end{equation}
where $H\equiv \dot{a}/a$ is the Hubble parameter and $k=4$. This energy density only mildly depends on $k$, with $\rho_{\rm end}\simeq (4.9\times 10^{15}\,{\rm GeV})^4$ for $k=10$.\par \medskip

\subsection{Coherent inflaton oscillations}\label{sec:sec22}

After the end of inflation the inflaton oscillates about the minimum of its potential. The dynamics can be described by the potential~(\ref{eq:Vphi}) which is a good approximation to Eq.~(\ref{eq:attractor}) already after one oscillation. Variation of the action~(\ref{eq:actionphi}) with respect to the homogeneous inflaton field and metric yields the equation of motion for the inflaton field and the Friedmann equation
\begin{align}\label{eq:eomback}
\ddot{\phi} + 3H\dot{\phi} + \lambda  k \phi^{k-1}\;&\simeq\;0\,,\\
\frac{1}{2}\dot{\phi}^2 + \lambda\phi^k\;&\simeq\; 3M_P^2H^2\,,
\end{align}
Multiplying (\ref{eq:eomback}) by $\phi$, and averaging over oscillations, it is straightforward to obtain the following virial-like relation for the mean kinetic energy, 
\begin{equation}
  \dfrac{1}{2} \langle \dot \phi^2\rangle \simeq  \dfrac{1}{2}  \langle \phi V_\phi \rangle \,,
\end{equation}
with $V_\phi \equiv \partial V/\partial \phi$. This allows us to express the oscillation-averaged energy and pressure densities as~\cite{Garcia:2020wiy}
\begin{align}\label{eq:rhoback}
\rho_{\phi} \;&\equiv\; \frac{1}{2}\langle \dot{\phi}^2\rangle + \langle V(\phi)\rangle \;\simeq\; \left( \dfrac{k+2}{2}\right)\langle V(\phi)\rangle \, ,\\
P_{\phi} \;&\equiv\; \frac{1}{2}\langle \dot{\phi}^2\rangle -  \langle V(\phi)\rangle \;\simeq\; \left( \dfrac{k-2}{2} \right) \langle V(\phi)\rangle \,.
\end{align}
The (oscillation-averaged) Equation-Of-State (EOS) parameter $w \equiv P_\phi/\rho_\phi$ can thus be uniquely determined in terms of the exponent $k$ through
\begin{equation}
    w \, = \, \dfrac{k-2}{k+2} \,.
    \label{eq:EOSk}
\end{equation}
This corresponds to a redshift of the energy density of $\phi$ of the form $\rho_{\phi}\simeq \rho_{\rm end}(a/a_{\rm end})^{-\frac{6k}{k+2}}$.

\begin{figure}
\centering{
\includegraphics[width=0.8 \linewidth]{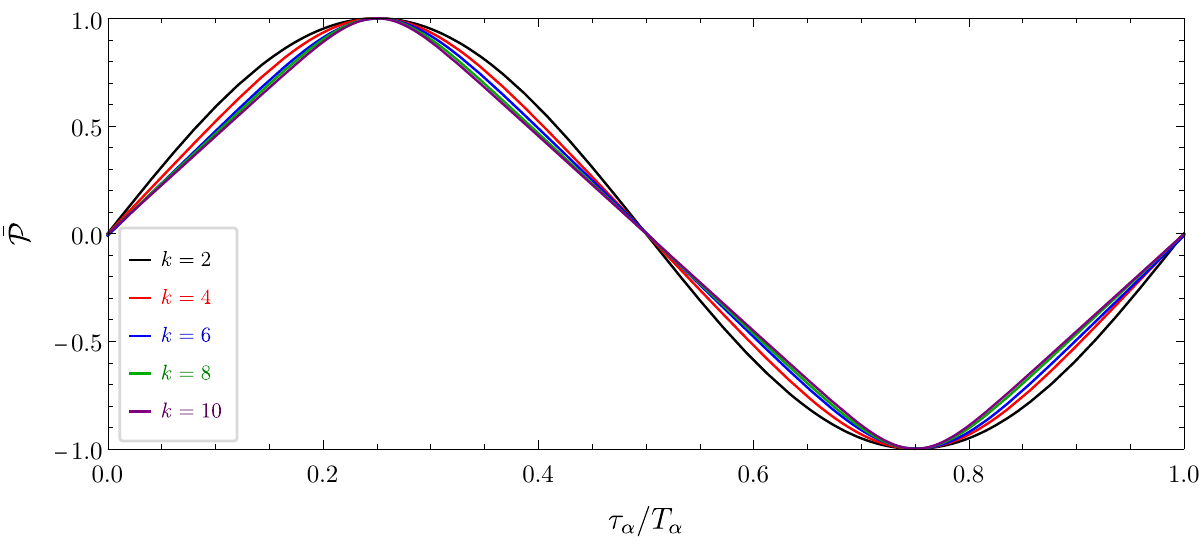}

}
\caption{Time evolution of the normalized periodic component of the inflaton field over an oscillation period.}\label{oscillationplot}
\end{figure}

During its oscillations about the minimum, the inflaton field can be expanded as
\begin{equation}
    \phi(t) \, = \, \phi_0 (t) \,\mathcal{P}(t) \,  \simeq \, \phi_\text{end} \left( \dfrac{a}{a_\text{end}} \right)^{-6/(k+2)}   \, \mathcal{P}(t)
\end{equation}
where $\phi_0 (t) \propto \rho_{\phi}^{1/k} \propto a^{-6/(k+2)}$ is a decaying envelope and $\mathcal{P}(t)$ is periodic function encapsulating the fast inflaton oscillations. To facilitate a $k$-universal treatment, it is convenient to define the rescaled periodic function $\bar{\mathcal{P}} \equiv (\phi_\text{end}/M_P)\mathcal{P} $ and to introduce a new $k$-dependent time variable related to cosmic time $t$ via $\diff \tau_\alpha \equiv (a/a_\text{end})^{-3(k-2)/k+2)} \sqrt{\lambda} M_P \diff t$ such that the EOM~(\ref{eq:eomback}) can be recast as
\begin{equation}\label{eq:taualpha}
    \dfrac{\diff^2 \bar{\mathcal{P}}}{\diff \tau_\alpha^2} + k \bar{\mathcal{P}}^{k-1} \, = \, 0 \,.
\end{equation}
Solutions to this equation can be found in implicit form as
\beq
\tau_{\alpha} \;=\; \frac{\bar{\mathcal{P}}}{\sqrt{2}} \, _2F_1\left(\frac{1}{2},\frac{1}{k};1+\frac{1}{k};\bar{\mathcal{P}}^{k}\right)\qquad \Rightarrow\qquad \bar{\mathcal{P}}(\tau_{\alpha}) \;=\; \begin{cases}
\sin(\sqrt{2}\tau_{\alpha})\,, & [k=2]\,,\\
{\rm sn}(\sqrt{2}\tau_{\alpha},-1)\,, & [k=4]\,,\\
\quad\vdots
\end{cases}
\label{eq:solEOMintermsoftaualpha}
\eeq
where ${}_2F_1$ are hypergeometric functions. The oscillating functions are represented in Fig.~\ref{oscillationplot} for $k=2,4,6,8,10$ as a function of the $\alpha$-time over one period of oscillation $T_\alpha$
\begin{equation}
    T_\alpha=2 \sqrt{2\pi} \left( \dfrac{\Gamma \left( 1+1/k \right)}{\Gamma \left( 1/2+1/k \right)} \right)
\end{equation}
where the initial time $\tau_\alpha=0$ as been chosen when $\bar{\mathcal{P}}$ is zero. Starting from a sine solution for $k=2$, by increasing $k$ higher harmonics of the periodic function $\bar{\mathcal{P}}$ contribute more significantly, driving the solution towards a ``triangular" periodic function.

\subsection{Growth of inhomogeneities}\label{sec:sec23}

Inhomogeneities in the post-inflation era are characterized by space and time dependent inflaton fluctuations $\delta\phi(t,\bx)$.~\footnote{In this work we ignore the growth of fluctuations of the metric, which can potentially be important during the preheating stage~\cite{Nambu:1996gf,Bassett:1998wg,Bassett:1999mt,Jedamzik:2010dq,Huang:2011gf,Giblin:2019nuv}.} By initially neglecting couplings of the inflaton to other degrees of freedom (corresponding to the term $\mathcal{L}_\text{int}$ in (\ref{eq:actionphi})), variation of the action leads to the equation of motion for the inflaton perturbation 
\beq\label{eq:eomdeltaphi}
\ddot{\delta\phi} + 3H\dot{\delta\phi} - \frac{\nabla^2 \delta\phi}{a^2} + k(k-1)\lambda \phi(t)^{k-2}\,\delta\phi \;=\; 0\,.
\eeq
One can rewrite the equation of motion in terms of the canonically normalized inflaton fluctuation $X=a\,\delta\phi$ expanded in Fourier space
\beq\label{eq:Xdef}
X(\tau,\bx) = \int \frac{\diff  ^3\bp}{(2\pi)^{3/2}}\,e^{-i\bp\cdot\bx} \left[ X_p(\tau)\hat{a}_{\bp} + X_p^*(\tau)\hat{a}^{\dagger}_{-\bp} \right]\,.
\eeq
where $\bp$ is the comoving momentum. The creation and annihilation operators $\hat{a}_{\bp}$ and $\hat{a}^{\dagger}_{\bp}$ satisfy the canonical commutation relations  $[\hat{a}_{\bp},\hat{a}_{\bp'} ] = [\hat{a}^{\dagger}_{\bp},\hat{a}^{\dagger}_{\bp'} ] = 0$ and $[\hat{a}_{\bp},\hat{a}^{\dagger}_{\bp'} ] = \delta(  \bp-\bp')$. In addition, the Wronskian constraint $
X_pX^{*\prime}_p - X_p^*X_p' \;=\; i$ is imposed to fulfill the canonical commutation relations between $X$ and its canonically conjugated momentum. Here ${}^{\prime}\equiv \diff/\diff\tau$ represents differentiation with respect to conformal time, $\diff \tau = \diff t/a$. Substitution of $\delta \phi = X /a $ into (\ref{eq:eomdeltaphi}) gives an equation describing the evolution of mode functions
\beq\label{eq:eomX}
X_{p}'' + \omega_p^2 X_{p} \;=\; 0\,,
\eeq
where the effective frequency reads
\beq\label{eq:omega}
\omega_p^2 \;\equiv\; p^2 - \frac{a''}{a} + k(k-1)\lambda a^2 M_P^2\left(\frac{\phi(t)}{M_P}\right)^{k-2}\,.
\eeq
To keep track of the growth of inhomogeneities during the post-inflation epoch, one solves the mode equation by assuming at some initial time $\tau_0$ the positive-frequency Bunch-Davies vacuum 
\beq
X_p(\tau_0) \;=\; \frac{1}{\sqrt{2\omega_p}} \,, \quad X_p'(\tau_0) \;=\; -\frac{i\omega_p}{\sqrt{2\omega_p}}\,,
\eeq
when the mode is deep inside the Hubble radius (i.e.~$p>a(\tau_0) H(\tau_0)$). The mode equation (\ref{eq:eomX}) features parametric instabilities driven by the oscillating term in the frequency~(\ref{eq:omega}) which can be made more explicit in terms of the periodic function $\bar{\mathcal{P}}$
\beq\label{eq:omega2}
\omega_p^2 \;\simeq \; p^2 + a^2 \Bigg[ \left( \dfrac{4-k}{2+k} \right) H_\text{end}^2  \left( \dfrac{a}{a_\text{end}} \right)^{-6k/(2+k)} + k(k-1)\lambda  M_P^2 \, \bar{\mathcal{P}}^{k-2} \, \left( \dfrac{a}{a_\text{end}} \right)^{-6(k-2)/(2+k)}   \Bigg] \,.
\eeq
were we used the second Friedmann equation $a''/a=(1-3w)\mathcal{H}/2$ with $\mathcal{H}\equiv a H$ the conformal Hubble parameter. Importantly, even for a weakly coupled theory with $\lambda \ll 1$, necessary for CMB compatibility c.f.~(\ref{eq:lambdapp}), parametric resonances ensure a significant growth of inhomogeneities if the oscillating term is active for a substantial amount of time. According to Floquet's theorem, the solution to the mode equation can be expressed as~\cite{Magnus2004-br}
\begin{equation}
X_p (\tau) \;=\; e^{\mu_p \tau}g_1(\tau) + e^{-\mu_p \tau}g_2(\tau)\,,    
\end{equation}
where $g_{1,2}$ are periodic functions and $\mu_p$ are complex numbers called the Floquet exponent. Eq.~(\ref{eq:eomX}) can be solved by identifying instability bands corresponding to $\text{Re}(\mu_p)>0$ signifying exponentially growing modes. We refer the reader to~\cite{Lozanov:2017hjm,Garcia:2023eol} for further details on the Floquet analysis. In general, instability bands are time-dependent through the scale factor dependence of the second terms in brakets in Eq.~(\ref{eq:omega2}). However, for $k=4$ all scale factor dependence cancels out as a consequence of conformal invariance of the quartic potential, as discussed in Ref.~\cite{Garcia:2023eol}. In our case, exponential growth will occur once modes cross the first narrow instability band~\cite{Lozanov:2017hjm}. One can estimate the typical scales subject to parametric resonances $p/(a_\text{end} \sqrt{\lambda} M_P) \simeq  \mathcal{O}(1), 1.1 \times 10^{-2},1.6 \times 10^{-5} , 7.6 \times 10^{-9} $ respectively for $k=4,6,8,10$~\cite{Garcia:2023dyf}. The growth of inhomogeneities eventually becomes large enough to trigger significant non-linearities and mode-mode couplings whose proper treatment requires numerical integration.

\subsection{Simulating fragmentation}\label{sec:sec24}

\begin{figure}
\centering{
\includegraphics[width=0.48 \linewidth]{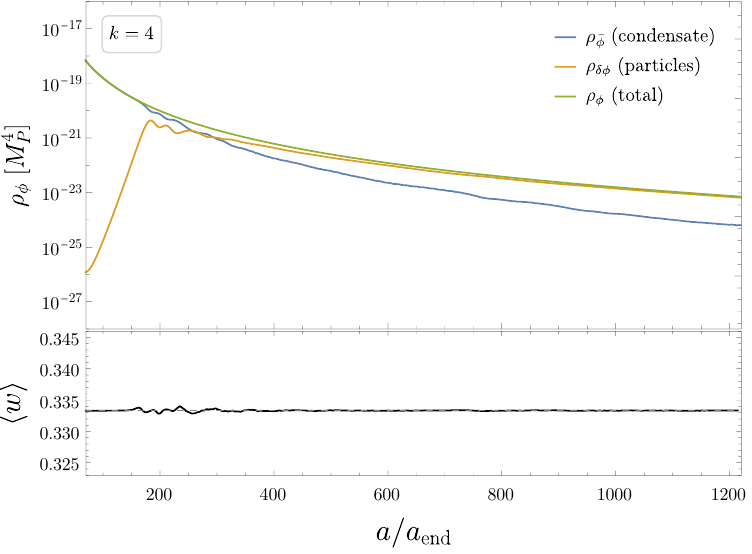}
\includegraphics[width=0.48 \linewidth]{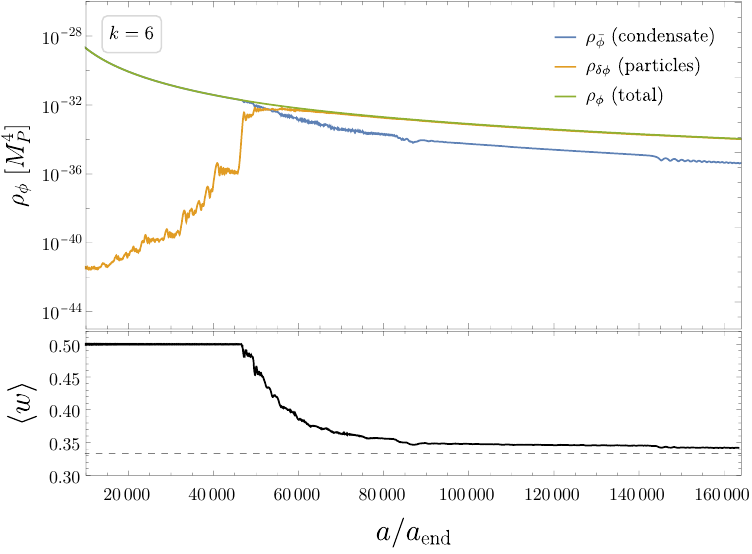}\\
\includegraphics[width=0.48 \linewidth]{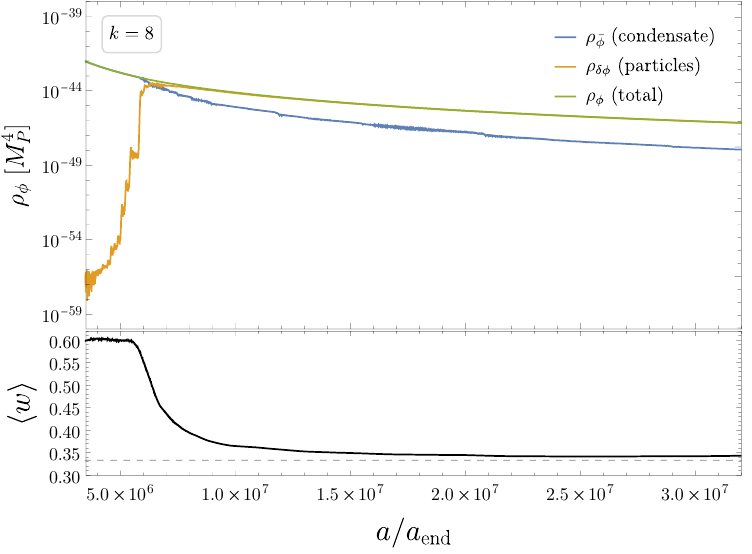}
\includegraphics[width=0.48 \linewidth]{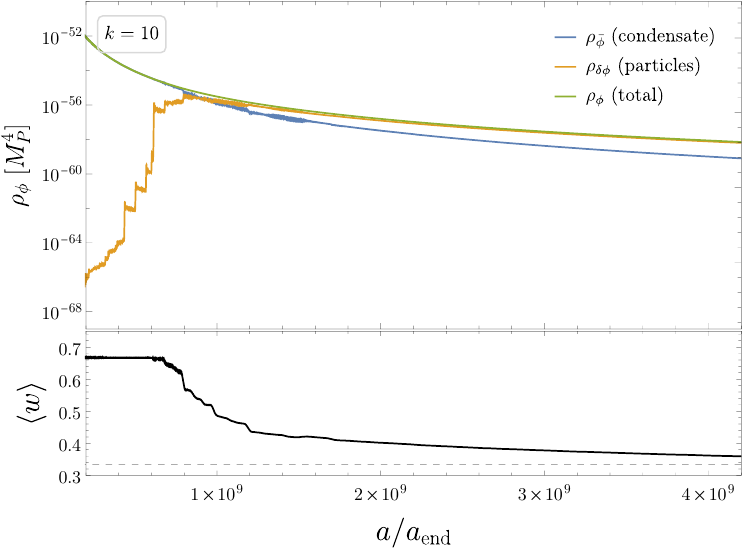}
}
\caption{Energy density of the inflaton in the form of condensate ($\rho_{\bar{\phi}}$, blue), in free quanta ($\rho_{\delta\phi}$, yellow) and the sum of both contributions ($\rho_{\phi}$, green) in the top panels. Oscillation-averaged equation of state in the bottom panels as a functions of the scale factor for $k=4,6,8,10$.}\label{energydensityplot}
\end{figure}

In order to account for non-linearities induced by the growth of inhomogeneities, we perform lattice simulations with the use of~\CL~\cite{Figueroa:2020rrl,Figueroa:2021yhd}. We first solve the background inflaton equation of motion during inflation and determine the value of various background parameters at the end of inflation. We use these values as initial conditions with~\CL~to simulate the post-inflation dynamics. The inflaton field is treated as a classical field on the lattice and initial conditions for inhomegeneities are randomly assigned to reproduce the statistical properties of the Bunch-Davies vacuum. The time variable used for the simulation is the $\alpha-$time $\tau_\alpha$ (see (\ref{eq:taualpha})) and the lattice parameters are chosen appropriately to fully capture parametric resonances. The total energy density of the inflaton is computed on the lattice from the spatially-averaged energy-momentum tensor of $\phi(\bx,\tau)$, denoted with an over-bar,
\beq
\rho_{\phi} \;=\; \overbar{\frac{1}{2}\dot{\phi}^2 + \frac{1}{2a^2}(\nabla\phi)^2+V(\phi)}\,.
\eeq
Following~\cite{Garcia:2023dyf,Garcia:2023eol}, one can define a condensate (i.e. homogeneous) component of the energy density as the sum of kinetic and potential energies of the spatially-averaged inflaton field~\cite{Garcia:2021iag}
\beq
\rho_{\bar{\phi}}  \;\equiv\; \frac{1}{2}\dot{\bar{\phi}}^{\,2} + V(\bar{\phi})\,.
\eeq
Inhomogeneities are characterized by the comoving occupation number of $\phi$, which in terms of mode functions and their time-derivatives is given by the adiabatic invariant~\cite{Kofman:1997yn}
\beq
n_{p} \;=\; \frac{1}{2\omega_p}\left| \omega_p X_p - i X'_p \right|^2\,.
\label{eq:PSDandoccupationnumber}
\eeq
By integrating this quantity, one can infer the number density and the energy density of the inflaton fluctuations expressed as the UV-regular forms~\cite{Kofman:1997yn,Garcia:2021iag}
\begin{align}\label{eq:ndeltaphi}
n_{\delta\phi} \;&=\; \frac{1}{(2\pi)^3a^3}\int \diff^3\bp\, n_p\,,\\ \label{eq:rhodeltaphi}
\rho_{\delta\phi} \;&=\; \frac{1}{(2\pi)^3a^4}\int \diff^3\bp\, \omega_p n_p\,.
\end{align}
Notably, the later expression corresponds to $\rho_{\delta\phi}=\rho_{\phi}-\rho_{\bar{\phi}}$~\cite{Garcia:2023eol,Garcia:2024rwg}. Each contribution to the energy-density resulting from our simulations is represented in Fig.~\ref{energydensityplot} as a function of the scale factor for $k=4,6,8,10$. The corresponding oscillation-averaged EOS parameter is in turn shown in the bottom panels. For all cases, initially the EOS is constant and follows the expected scaling relation of Eq.~(\ref{eq:EOSk}) as the universe is dominated by the coherently oscillating condensate. The growth of inhomogeneities can be observed from these plots as a sharp increase in $\rho_{\delta \phi}$ until backreaction is reached and the inflaton condensate is fragmented.  We estimate that fragmentation is reached at $a_\text{frag}/a_\text{end} \simeq 180,4.5 \times 10^4,6 \times 10^6,7\times 10^8$ for $k=4,6,8,10$ consistently with Ref.~\cite{Lozanov:2016hid,Lozanov:2017hjm,Garcia:2023dyf,Garcia:2023eol}. Following this event, the condensate component redshifts away and  becomes subdominant as the universe become filled with a collection of relativistic inflaton quanta dominating the energy budget. The EOS relaxes progressively to 1/3 over typical time scales of $\mathcal{O}(2-3)(a_\text{frag}/a_\text{end})$, slightly dependent on the value of $k$.

\subsection{Transition to an inflaton-quanta dominated phase} \label{sec:sec25}

\begin{figure}[t]
\centering{
\includegraphics[width=0.43 \linewidth]{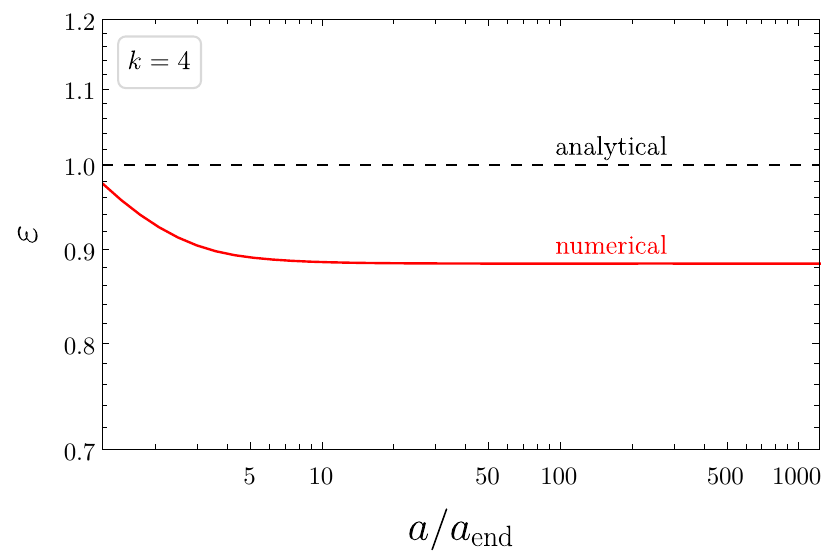}
\includegraphics[width=0.43 \linewidth]{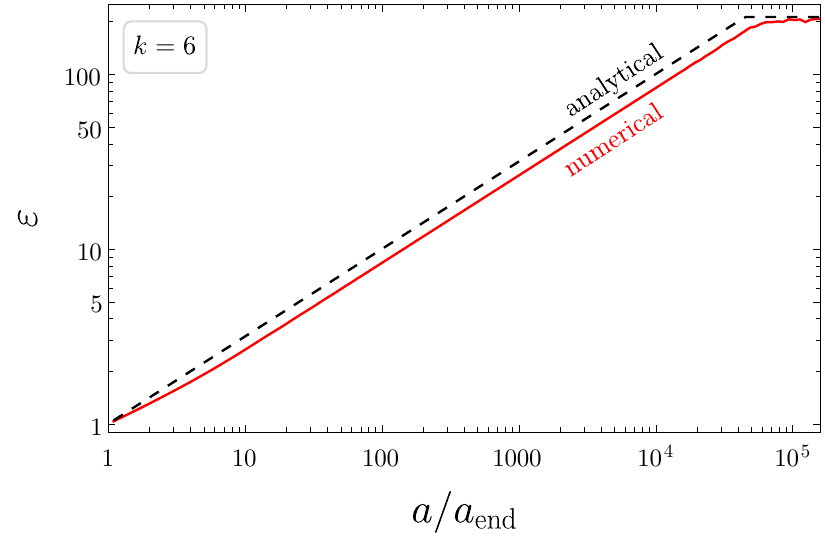}\\
\includegraphics[width=0.43 \linewidth]{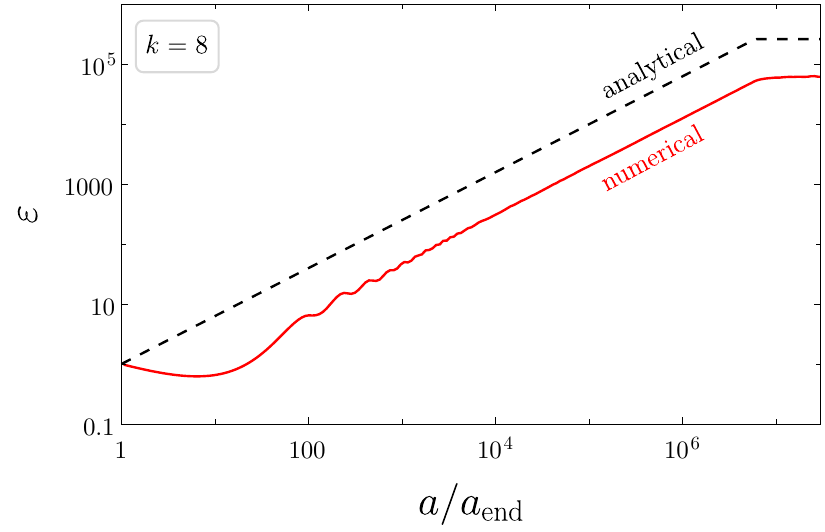}
\includegraphics[width=0.43 \linewidth]{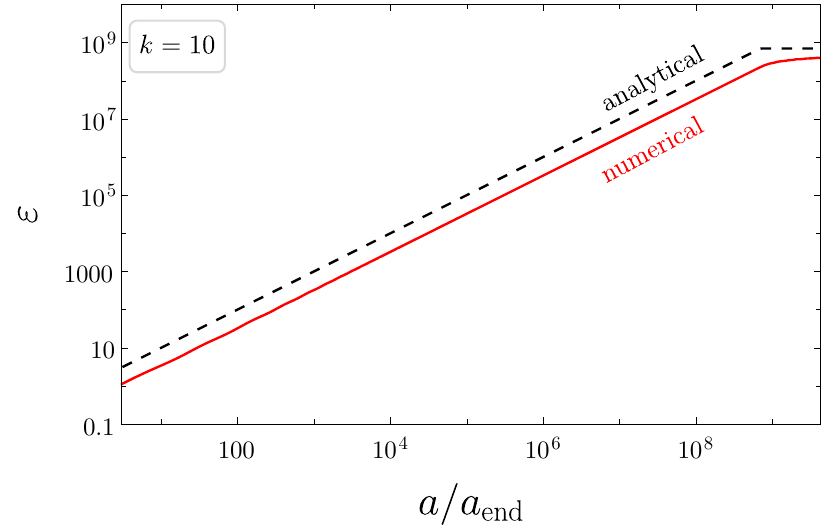}
}
\caption{ Analytical (black dashed lines) and numerical (red lines) evaluation of the redshift parameter $\varepsilon$ defined in Eq.~(\ref{eq:varepsilon}) as a function of the scale factor for $k=4,6,8,10$.}\label{epsilonplot}
\end{figure}

\noindent
As it can be appreciated in Fig.~\ref{energydensityplot}, the transition from the condensate EOS to a radiation dominated universe is not instantaneous. For this reason, it is convenient to introduce the scale factor $a_\text{RD}>a_\text{frag}$ at which the universe is almost exclusively composed of a population of fragmented relativistic inflaton particles. That is, $\frac{k-2}{k+2}>w>1/3$ for $a_{\rm frag}<a<a_{\rm RD}$, and $w\simeq 1/3$ for $a>a_{\rm RD}$. In practice we take $a_\text{RD}$ as the scale factor evaluated at the end of our simulations. The expansion dynamics of the universe during the post-inflationary stage can be parameterized by the time-dependence of its equation-of-state $w(a)$. Straightforward integration of the continuity equation $\dot \rho + 3 H \rho (1+w) = 0$  allows to express the total energy-density $\rho$ at a given time as
\begin{equation}
 \rho (a) \, = \, \rho_\text{end} \left( \dfrac{a}{a_\text{end}}\right)^{-3} \exp \left[ -3 \int_{a_\text{end}}^{a} w(\tilde a) \diff \log \tilde a\right] \,.
\end{equation}
  We now introduce a redshift factor accounting for time-dependent deviations with respect to a strict radiation-like (i.e. $w=1/3$) dominated universe~\cite{Cosme:2022htl}
\begin{equation}
    \varepsilon(a) \, \equiv \,   \left( \dfrac{a}{a_\text{end}} \right)^{3\bar{w}(a) -1} \,,
    \label{eq:varepsilon}
\end{equation}
defined in terms of the instantaneous (logarithmically) averaged equation-of-state
\begin{equation}
    \bar{w}(a)  \, \equiv \, \dfrac{1}{\log(a/a_\text{end})} \int_{a_\text{end}}^{a} w(\tilde a) \diff \log \tilde a \,.
\end{equation}
By construction, values of $ \varepsilon (a) $ are larger (smaller) than unity for an EOS $w>1/3$ ($w<1/3$). The redshift factor $\varepsilon (a) $ also tends to a constant value after fragmentation, for $a \geq a_{\rm RD} \gg a_\text{frag}$, as the EOS asymptotes 1/3. The numerical evaluation of this factor from our lattice simulations is represented in Fig.~\ref{epsilonplot}, showing how $\varepsilon(a)$ exhibits an asymptotic value reached at late time as the oscillation-averaged EOS of Fig.~\ref{energydensityplot} asymptotes 1/3. 

In what follows, abusing somewhat notation, we will refer to the asymptotic value of the redshift factor by simply dropping the argument, $ \varepsilon \equiv \varepsilon (a \geq a_\text{RD}) $. In terms of it we can express the scale factor during the radiation-like dominated era as
\begin{equation}
 \left( \dfrac{a_\text{end}}{a} \right) \, = \, \left( \dfrac{\rho(a)}{\rho_\text{end}}\right)^{1/4} \, \varepsilon^{1/4}~~\qquad  [a\geq a_\text{RD}] \,,
\end{equation}
which holds in particular at reheating $a=a_\text{reh}$. This allows us to express the scale factor at the end of inflation as 
\begin{equation}
 \left( \dfrac{a_\text{end}}{a_0} \right)  \, = \, \varepsilon^{1/4} \, \left( \dfrac{\rho_{\text{rad},0}}{\rho_\text{end}} \right)^{1/4} \left( \dfrac{g_\text{reh}}{g_0} \right)^{1/4} \left( \dfrac{g_{s,0}}{g_{s,\text{reh}}} \right)^{1/3}  \, ,
\label{eq:scalefactorend}
\end{equation}
where we have used entropy conservation, $\rho_\text{reh}^{1/4}a_\text{reh} = \rho_{\text{rad},0}^{1/4}a_0 (g_\text{reh}/g_0)^{1/4} (g_{s,0}/g_{s,\text{reh}})^{1/3}$. Here $\rho_{\text{rad},0}$ denotes the radiation energy density at the present day, and $g_{\rho}$ and $g_{s}$ are the energy density and entropy degrees of freedom, respectively, which have been evaluated at reheating and present day.\par \medskip

A simple analytical estimate of the parameter $\varepsilon$ can be obtained by assuming instantaneous fragmentation and approximating the EOS parameter posterior to inflation by a step function
\begin{equation}
     w_\text{approx}(a) \, \simeq \, \left( \dfrac{k-2}{k+2} \right)\Theta \big( a_\text{frag}-a \big)+ \dfrac{1}{3}\Theta \big( a-a_\text{frag} \big)\,,
     \label{eq:wapprox}
\end{equation}
with $\Theta $ being the Heaviside function. This immediately yields
\begin{equation}
    \varepsilon_\text{approx} \, \simeq \,  \left(\dfrac{ a_\text{frag}}{a_\text{end}}\right)^{2(k-4)/(2+k)} \,\simeq \begin{cases}
1 &  [k=4]\,,\\
2.1 \times 10^2 \, \left(\dfrac{ a_\text{frag}/a_\text{end}}{4.5 \times 10^4}\right)^{1/2} ~~~~ & [k=6]\,,\\
2.6 \times 10^5 \, \left(\dfrac{ a_\text{frag}/a_\text{end}}{6 \times 10^6}\right)^{4/5}  \qquad \quad  & [k=8]\,,\\
7 \times 10^8\, \left(\dfrac{ a_\text{frag}/a_\text{end}}{7 \times 10^8}\right) &  [k=10]\,.
\end{cases}
\end{equation}
in good agreement with numerical results $\varepsilon_\text{num}\simeq \{0.88,208,6.3 \times 10^4,4 \times 10^8\}$ respectively for $k=\{4,6,8,10\}$ represented in Fig.~\ref{epsilonplot}. In this same Figure we show for comparison the analytical estimates of $\varepsilon(a)$ based on Eq.~(\ref{eq:wapprox}). The small deviation from $\varepsilon=1$ for $k=4$ comes from the fact that the inflation potential slightly departs, in the first post-inflation instants, from an exactly quartic potential. Similar effects are also relevant for larger $k$ but less significant as $\varepsilon \gg 1$. Our numerical treatment allow us to fully capture these effects. \par \medskip


\subsection{Reheating} \label{sec:sec26}

After inflation ends, the energy density of the inflaton would have been transferred to relativistic degrees of freedom via the reheating mechanism. Reheating can be achieved by coupling the inflaton to other species that could be fermions ($f$) or bosons ($b$) and that could belong to (or could be connected to) the Standard-Model thermal plasma. For simplicity, in this work we consider the energy transfer from the inflaton to a relativistic species without specifying their interaction properties with respect to the Standard-Model. Such energy transfer could be induced  via renormalizable interaction terms in the Lagrangian such as 
\beq
\mathcal{L}_\text{int}\supset  \begin{cases} 
y \phi \bar{f}f  \\
\mu \phi bb \\
\sigma \phi^2b^2,
\end{cases}
\label{couplings}
\eeq
where $y, \sigma$ ($\mu$) are dimensionless (dimensionful) parameters. 
At the early stages of reheating, prior to fragmentation, the particle production process is dominated by the transition from the homogeneous oscillating condensate $\phi(t)$ to the final-state free particles. Disregarding collective enhancement/blocking effects (preheating), which can be done for sufficiently small Lagrangian couplings, the interactions (\ref{couplings}) induce a source term for the energy density of the radiation component ($R$) of the form~\cite{Garcia:2020wiy,Garcia:2023dyf}
\beq
\dot{\rho}_{R} + 4H \rho_{R} \;\simeq\; (1+w)\Gamma_{\phi}\rho_{\bar{\phi}}\, , 
\label{eq:aeom}
\eeq
in addition to modifying the Friedmann equation
\beq
\rho_{\bar{\phi}}+\rho_{R} \;=\; 3H^2 M_P^2\,.
\label{hub}
\eeq

The corresponding energy-transfer rate from the inflaton condensate into bosonic or fermionic final states can be expressed as
\beq
\Gamma_\phi = \begin{cases} 
 \dfrac{y_{\rm eff}^2(k)}{8\pi}  m_{\phi}(t)\,,  \\[8pt]
\dfrac{\mu_{\rm eff}^2(k)}{8\pi  m_{\phi}(t) }\,,  \\[8pt]
\dfrac{\sigma_{\rm eff}^2}{8 \pi} 
\dfrac{\rho_\phi(t)}{ m^3_\phi(t)}\,,
\end{cases}
\label{couplings2}
\eeq
with the time-dependent inflaton mass given by $m_\phi(t)\simeq \sqrt{k(k-1)}\lambda^{1/k}\rho_\phi^{(k-2)/2k}$.  The effective couplings appearing in Eq.~(\ref{couplings2}) can be expressed for fermion decay products as 
\begin{equation}
    y_{\rm eff}^2(k) \;\equiv\;  \sqrt{\frac{\pi k}{2(k-1)}}\,\left( \frac{\Gamma(\frac{1}{2}+\frac{1}{k})}{\Gamma(\frac{1}{k})} \right) \alpha_y(k,\mathcal{R}) y^2\, .
\end{equation}
Similarly, bosonic effective couplings can be expressed as~\cite{Garcia:2020wiy}
 \beq 
\mu_{\rm eff}^2(k) 
\equiv\; \frac{1}{4}(k+2)(k-1) \sqrt{\frac{\pi k}{2(k-1)}}\,\left(\frac{\Gamma(\frac{1}{2}+\frac{1}{k})}{\Gamma(\frac{1}{k})}\right) \alpha_{\mu}(k,\mathcal{R}) \mu^2\,,
\label{eq:zeffdef}
\eeq 
and
\beq
\sigma^2_{\rm eff}(k) 
\equiv\; \frac{1}{8} k(k+2)(k-1)^2 \sqrt{\frac{\pi k}{2(k-1)}}\,\left(\frac{\Gamma(\frac{1}{2}+\frac{1}{k})}{\Gamma(\frac{1}{k})} \right) \alpha_{\sigma}(k,\mathcal{R}) \sigma^2\,.
\label{eq:seffdef}
\eeq
These effective couplings depend on the details of the oscillations throughout the exponent $k$. In addition, they involve the efficiency parameters $\alpha_y(k,\mathcal{R})$,  $\alpha_{\mu}(k,\mathcal{R})$ and $\alpha_{\sigma}(k,\mathcal{R}) $ which account for the kinematic effects resulting from effective masses $m_{\rm eff}$ of the decay products induced by the time-dependent inflaton field vev, $\mathcal{R}\propto(m_{\rm eff}/m_{\phi})^2_{\phi\rightarrow\phi_0}$. Details can be found in Ref.~\cite{Garcia:2020wiy}. \par \medskip

If the decay of the inflaton field is not completed prior to fragmentation, then the completion of the reheating process relies on the decay of the free inflaton particles (fluctuations), which now dominate the energy budget of the universe, into Standard-Model fields. In this case, the continuity equation (\ref{eq:aeom}) is modified to~\cite{Garcia:2023eol,Garcia:2023dyf,Garcia:2024rwg}
\beq\label{eq:rhorfrag}
\dot{\rho}_R + 4H\rho_R \;=\; \Gamma_{\delta\phi} m_{\phi} n_{\delta\phi}\,,
\eeq
where $n_{\delta\phi}$ denotes the free inflaton number density, and $\Gamma_{\delta\phi}$ is the free inflaton particle decay rate. To obtain an exact solution, the evaluation of the source term of (\ref{eq:rhorfrag}) requires the numerical determination of the inflaton occupation numbers from the lattice code~\cite{Garcia:2023eol,Garcia:2024rwg}. Nevertheless, this source term can be approximated upon the estimation of the non-relativistic particle energy density in terms of the corresponding Lorentz boost factor,
\beq
m_{\phi} n_{\delta\phi} \;\approx\; \left(\frac{m_{\phi}}{\bar{E}}\right)\rho_{\delta\phi}\qquad \text{with}\qquad \bar{E} \;=\; \left( \frac{\pi^{7/2}}{30^{3/4}\zeta(3)} \right)\rho_{\delta\phi}^{1/4}\,.
\eeq
The full expressions for the particle production rates can be found in~\cite{Garcia:2023dyf}.

The reheating time $a_\text{reh}$ can be defined by the condition $\rho_R(a_\text{reh}) = \rho_\phi(a_\text{reh})$ where $\rho_\phi$ accounts for both contributions from the condensate and the fragmented quanta. 
The conditions for reheating to occur after fragmentation, i.e. $a_\text{reh}>a_\text{frag}$, were derived in Ref.~\cite{Garcia:2023dyf} and are summarized here.\par \medskip

\noindent
For fermionic final states, requiring that reheating is achieved after fragmentation is equivalent to the condition
\beq
y_{\rm eff} < \left[ \frac{8 \pi (7-k)}{\sqrt{3} \lambda^\frac1k k \sqrt{k(k-1)}} \left(\frac{\rho_{\rm end}}{M_P^4} \right)^\frac1k \right]^{\frac12} \left( \dfrac{a_\text{frag}}{a_\text{end}} \right)^{-\frac{3}{k+2}}\,\qquad [8+k-6kl>0]\, ,
\label{ylim}
\eeq
where we introduced the quantity
\beq
l \;\equiv\; 
\begin{cases}
\frac{1}{2}-\frac{1}{k}\,,\quad & \phi\rightarrow \bar{f}f\,,\\
\frac{1}{k}-\frac{1}{2}\,,\quad & \phi\rightarrow bb\,,\\
\frac{3}{k}-\frac{1}{2}\,,\quad & \phi\phi\rightarrow bb\, ,
\end{cases}
\label{ls}
\eeq
parametrizing the power of $m_\phi$ appearing in the production rates. For  $8+k-6kl<0$, this condition becomes
\beq
y_{\rm eff} < \left[ \frac{8 \pi (k-7)}{\sqrt{3} \lambda^\frac1k k \sqrt{k(k-1)}} \left(\frac{\rho_{\rm end}}{M_P^4} \right)^\frac1k \right]^{\frac12} \left( \dfrac{a_\text{frag}}{a_\text{end}} \right)^{-\frac{k-4}{k+2}} \,\qquad [8+k-6kl<0]\, ,
\label{ylim-}
\eeq
In the case where inflaton decay products are scalars, the post-fragmentation reheating condition imply
\beq
\mu_{\rm eff} < \left[ \frac{8 \pi (2k+1)\sqrt{k(k-1)}\lambda^\frac1k M_P^2}{\sqrt{3} k  } \left(\frac{\rho_{\rm end}}{M_P^4} \right)^{(1-\frac1k)} \right]^\frac12 \left( \dfrac{a_\text{frag}}{a_\text{end}} \right)^{-\frac{3k-3}{k+2}} \, ,
\label{mlims}
\eeq
for the dimensionful coupling and 
\beq
\sigma_{\rm eff} < \left[ \frac{8 \pi (2k-5)(k(k-1))^\frac32 \lambda^\frac3k }{\sqrt{3} k  } \left(\frac{\rho_{\rm end}}{M_P^4} \right)^{(1-\frac3k)} \right]^\frac12 \left( \dfrac{a_\text{frag}}{a_\text{end}} \right)^{-\frac{3k-9}{k+2}} \, .
\label{slimsc}
\eeq
for the dimensionless coupling. An important point to emphasize is that since  the equation of state $w=1/3$ is maintained throughout the transition between the quanta dominated universe and the thermal plasma era, the precise moment of reheating does not play any role on relevant physical observables. Particularly, as long as the previously stated conditions are satisfied, the evolution of the scale factor would remain independent on the inflaton couplings to the radiation bath and the gravitational wave signal computed in the following section would remain unaffected.\footnote{However, this coupling must be sufficiently large to ensure reheating to be achieved before nucleosynthesis.} 

\par \medskip

\begin{figure}[t]
\centering{
\includegraphics[width=0.7 \linewidth]{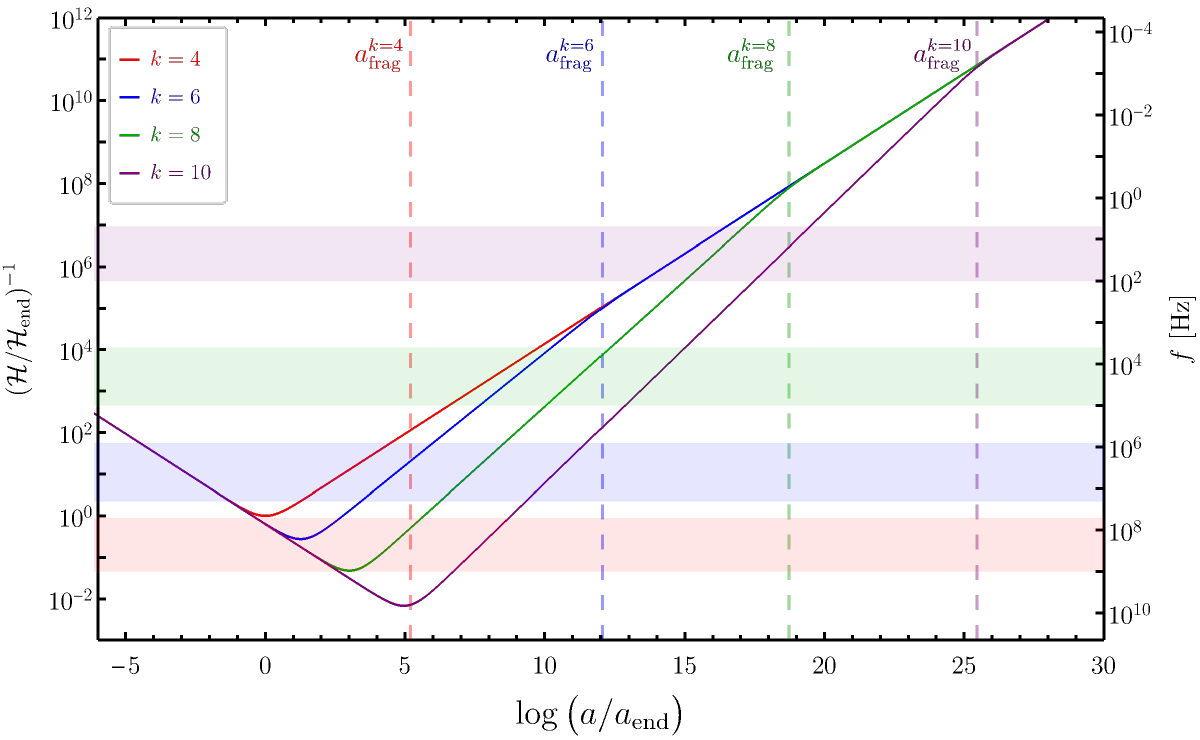}
}
\caption{Comoving Hubble radius and relevant scales normalized to the value at the end of inflation $\mathcal{H}_\text{end}^{-1}$ at $a=a_\text{end}$ for $k=4$. The corresponding gravitational wave frequency at the present day is represented on the vertical axis on the right. Colored bands represent the scales prone to parametric resonances that are typically excited during the fragmentation process (see (\ref{eq:freqs})). }\label{scales}
\end{figure}

\noindent
\textbf{Summary.} The conditions (\ref{ylim-})-(\ref{slimsc}) ensure that reheating is achieved after the fragmentation process is completed and that the equation of state transits smoothly from $w=(k-2)/(k+2)$ (where the conformal Hubble rate scales as $\mathcal{H}^{-1} \sim a^{2(k-1)/(k+2)}$) during the coherently-oscillating phase to $w=1/3$ (with $\mathcal{H}^{-1} \sim a$) during subsequent phases of the universe dominated by fragmented inflaton quanta or ultra-relativistic particles belonging to the thermal plasma. This completes the transition from cosmic inflation (with $\mathcal{H}^{-1} \sim a^{-1}$) to the standard phase of radiation domination. The cosmological history of our setup in summarized in Fig.~\ref{scales}, where the evolution of the comoving Hubble horizon $\mathcal{H}^{-1}$ is presented as a function of the scale factor throughout the various phases for each value of $k=4,6,8,10$ considered in this work.

\section{Gravitational waves}\label{sec:gws}

In this section we detail the computation of the gravitational wave signal generated throughout the entire cosmological history, accounting for the transition from cosmic inflation to a universe dominated by a coherently oscillating condensate followed by the domination of inflaton quanta and finally a thermal bath of relativistic particles. Limiting ourselves to tensor metric fluctuations, the perturbed Friedmann-Robertson-Walker metric can be parameterized by
\begin{equation}
    \diff s^2\,=\,a(\tau)^2\Big[ \diff \tau^2 - \Big( \delta_{ij} + h_{ij} \Big) \diff x^i \diff x^j   \Big]\,.
\end{equation} 
Gravitational waves are described by the metric perturbation $h_{ij}$ which represents two independent tensor degrees of freedom satisfying the transverseness and tracelessness (TT) conditions $\partial_i h_{ij}=0$ and $h^i_i=0$. In cosmological perturbation theory, tensor metric perturbations are decoupled from scalar perturbations at first order. Scalar perturbations only appear as a source of tensor metric perturbations at second order through the transverse-traceless component of the anisotropic stress $\Pi^\text{TT}_{ ij } = \big[\partial_i \phi \partial_j \phi \big]^\text{TT}$. Accounting for this source term, the corresponding equation of motion for tensor modes expressed in Fourier space reads
\begin{equation}
h^{\prime \prime}_{ij}(\bp,\tau)+ 2 \mathcal{H} h^{\prime }_{ij}(\bp,\tau)+p^2 h_{ij}(\bp,\tau) \, = \, \dfrac{2 a^2}{M_P^2} \Pi^\text{TT}_{ ij } (\bp,\tau)\,,
\label{eq:tensorEOM}
\end{equation}	
where $\bp$ is the (comoving) Fourier scale, $p=|\bp|$ and primes denote differentiation with respect to conformal time. As Eq.~(\ref{eq:tensorEOM}) is linear in $h_{ij}$, its solutions can be decomposed in terms of the sum of homogeneous and inhomogeneous solutions. 
During singe-field slow-roll inflation the source term is essentially negligible making the inhomogeneous solution irrelevant. For modes with wavelengths smaller than the Hubble radius at all times ($p>\mathcal{H}$), the homogeneous solution to this equation describes quantum fluctuations in the Bunch Davies vacuum and do not contribute to the Gravitational Wave (GW) signal. Modes experiencing a super-horizon phase ($p<\mathcal{H}$) during inflation would be excited, freezing-out outside the horizon, and contributing to the homogeneous solution if they remain outside the horizon, or sourcing the $B$-mode polarization spectrum of the CMB if they re-enter the horizon during the recombination epoch. The inhomogeneous solution is only significant after the end of inflation, once non-linearities induce sizable mode-mode couplings resulting in a non-vanishing anisotropic stress, efficiently sourcing gravitational waves during the fragmentation process. This section is dedicated to estimating contributions coming from the inhomegenous solution (i.e. induced by the fragmentation process) and the homogenous solution (i.e. tensor modes excited during inflation and re-entering during a stiff EOS era) and reconstructing the total gravitational wave signal.


\subsection{Gravitational waves from fragmentation} \label{sec:gws1}

As mentioned previously both contributions to the GW signal can be estimated separately, in this subsection we estimate the GW signal generated solely during the fragmentation process. At the onset of fragmentation, non-linearities become large enough to contribute significantly to gravitational wave production by acting as a source term through the anisotropic stress in the right-hand side of the equation of motion~(\ref{eq:tensorEOM}). Projected onto the transverse-traceless component in Fourier space, the anisotropic stress  can be written explicitly as~\cite{Dufaux:2007pt}
\begin{equation}
\Pi^\text{TT}_{ ij } (\bp,\tau) \, = \left( \, P_{i\ell}(\hat \bp) P_{jm}(\hat \bp)-\dfrac{1}{2} P_{ij}(\hat \bp) P_{\ell m}(\hat \bp) \right) \Pi_{ \ell m } (\bp,\tau)\,,
\end{equation}
with $P_{ij}=\delta_{ij}-\hat{p_i}\hat{p_j}$, $\hat p_i=p_i/p$ and
\begin{equation}
     \Pi_{ \ell m } (\bp,\tau) \, = \, \int\dfrac{\diff^3 \bq }{(2 \pi)^{3/2}} \, q_\ell \, q_m \, \phi(\bq,\tau) \, \phi(\bp-\bq,\tau) \,.
\end{equation}
We simulate numerically the production of gravitational waves with \CL~which implements a discretized TT projector.~\footnote{Such a discrete projector is not unique and might result in differences that were found to only marginally affect the UV part of the GW spectrum~\cite{Figueroa:2011ye,Figueroa:2020rrl}. } 
\par \medskip

\subsubsection{Comoving scales and frequencies}
 The gravitational wave frequency $f$ at the present epoch can be related to the comoving Fourier scale $p$ via~\cite{Dufaux:2007pt} 
\begin{equation}
f \,= \, \dfrac{p}{2 \pi a_0} \, = \, \varepsilon^{1/4} \, \left( \dfrac{\rho_{\text{rad},0}}{\rho_\text{end}} \right)^{1/4} \left( \dfrac{g_\text{reh}}{g_0} \right)^{1/4} \left( \dfrac{g_{s,0}}{g_{s,\text{reh}}} \right)^{1/3}  \dfrac{\kappa}{2 \pi} \, \sqrt{\lambda} \, M_P  \, ,
\label{eq:translatektof}
\end{equation}
where we introduced the dimensionless comoving momentum $\kappa  \equiv p/( a_\text{end} \sqrt{\lambda} M_P )$. An evaluation of this expression gives the following typical frequencies 
\begin{equation} \label{eq:freqs}
    f \, \simeq \, \begin{cases}
1.41 \times 10^8 \, \left( \dfrac{\kappa}{4} \right)~\text{Hz}  & [k=4],\\
8.6 \times 10^5 \, \left(\dfrac{ a_\text{frag}/a_\text{end}}{4.5 \times 10^4}\right)^{1/8} \left( \dfrac{\kappa}{1.5 \times 10^{-2}} \right) ~\text{Hz} & [k=6]\,,\\
2.05 \times 10^3 \, \left(\dfrac{ a_\text{frag}/a_\text{end}}{6 \times 10^6}\right)^{1/5}   \left( \dfrac{\kappa}{2.1 \times 10^{-5}} \right) ~\text{Hz} & [k=8]\,,\\
3.8 \, \left(\dfrac{ a_\text{frag}/a_\text{end}}{7 \times 10^8}\right)^{1/4} \left( \dfrac{\kappa}{9.7 \times 10^{-9}} \right) ~\text{Hz}  & [k=10]\,,
\end{cases}
\end{equation}
where normalization for $\kappa$ has been appropriately chosen as the typical scales prone to parametric resonance. A range of scales close to the typical $\kappa$ is represented as colored bands in Fig.~\ref{scales}. We denote the largest Fourier scale to leave the horizon during inflation by $p_\text{end}\equiv \mathcal{H}_\text{end}$, and the scale which re-enters the horizon at fragmentation by $p_\text{frag}\equiv  \mathcal{H}_\text{frag}$. Both scales represent respectively the scale tangent to the minimum of $\mathcal{H}^{-1}$ and the scale at which the vertical line corresponding to $a_\text{frag}$ and $\mathcal{H}^{-1}$ meet in Fig.~\ref{scales}, for each value of the exponent $k$. Under the approximation of instantaneous fragmentation, the corresponding frequencies can be related via
\begin{equation}
    f_\text{frag} \, \simeq \, f_\text{end} \, \left( \dfrac{a_\text{end}}{a_\text{frag}} \right)\, \varepsilon^{-1/2}
    \label{eq:pfrag}
\end{equation}
where $f_\text{end}$ can be expressed in term of the energy density at the end of inflation through $a_\text{end}$ via Eq.~(\ref{eq:scalefactorend}). We find
\begin{equation}
  f_\text{frag} \, \simeq \, \begin{cases}
2.6\times 10^5~\text{Hz}  & \\
2.7 \times 10^2~\text{Hz} & \\
4.8 \times 10^{-1}~\text{Hz} & \\
4.7 \times 10^{-4}~\text{Hz}  & 
\end{cases}  \,, \qquad \text{and} \qquad f_\text{end} \, \simeq \, \begin{cases}
4.5 \times 10^7~\text{Hz}  & \qquad [k=4],\\
1.7 \times 10^8~\text{Hz} & \qquad [k=6],\\
7.3 \times 10^8~\text{Hz} & \qquad [k=8],\\
6.5 \times 10^9~\text{Hz}  & \qquad [k=10].
\end{cases}
\label{eq:ffragandfend}
\end{equation}

\subsubsection{Energy density and spectra}

\begin{figure}[t]
\centering{
\includegraphics[width=0.40 \linewidth]{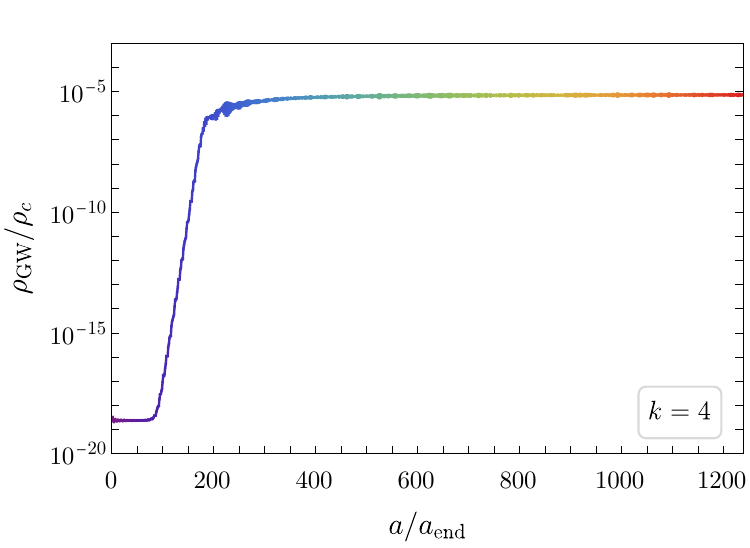}
\includegraphics[width=0.40 \linewidth]{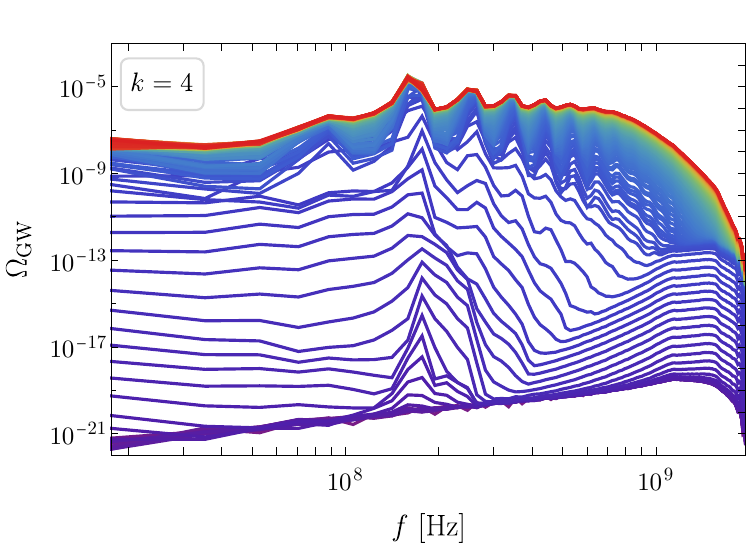}\\
\includegraphics[width=0.40 \linewidth]{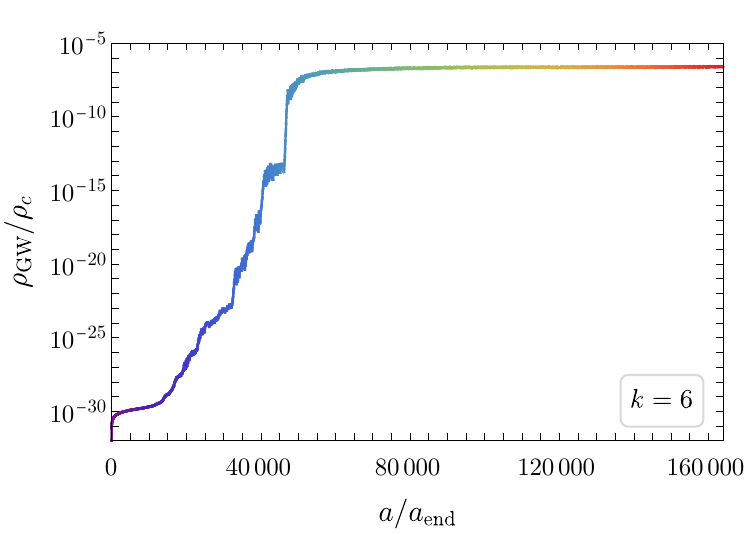}
\includegraphics[width=0.40 \linewidth]{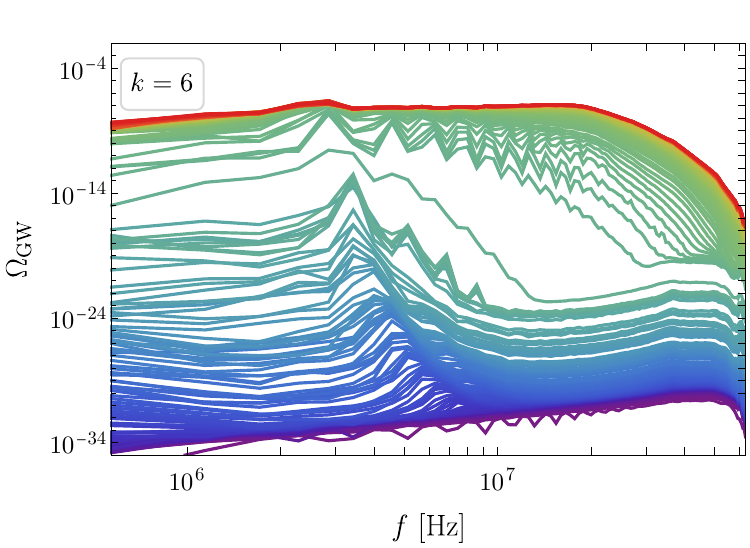}\\
\includegraphics[width=0.40 \linewidth]{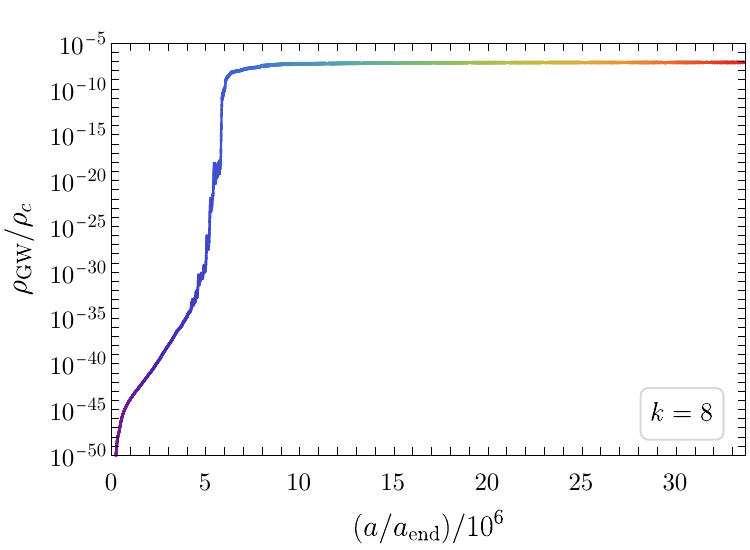}
\includegraphics[width=0.40 \linewidth]{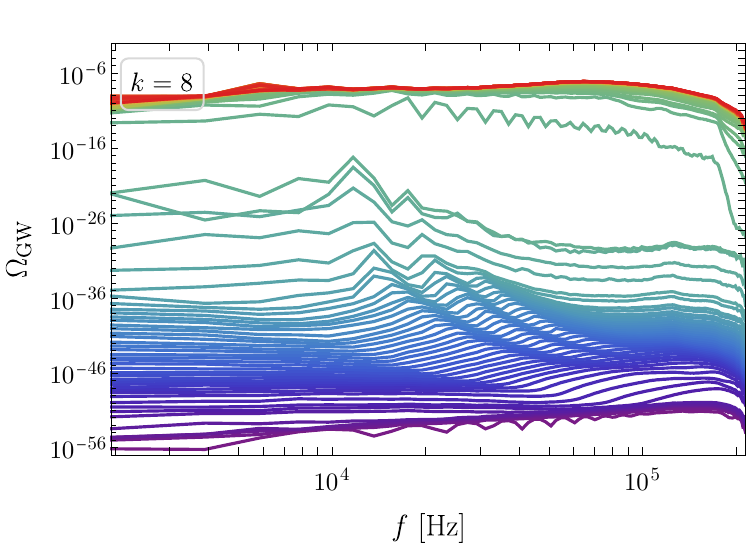}\\
\includegraphics[width=0.40 \linewidth]{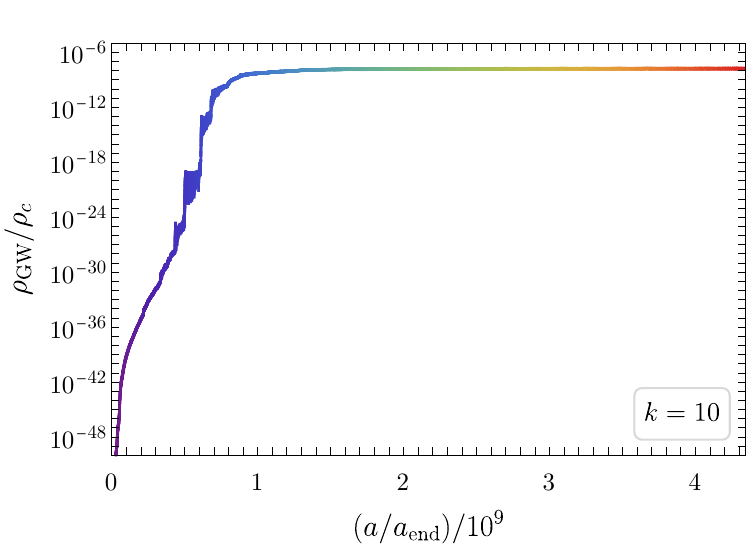}
\includegraphics[width=0.40 \linewidth]{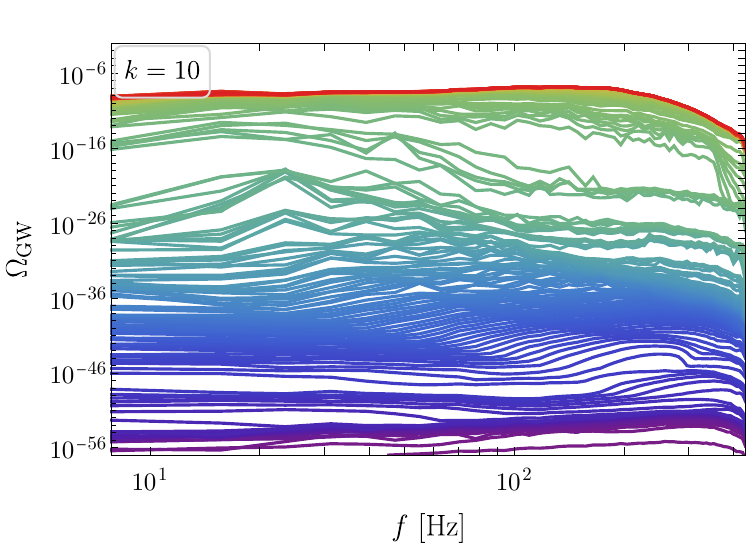}
}
\caption{Normalized gravitational wave energy density generated solely during the fragmentation process: as a function of the scale factor (left panel) and as a function of the present-day frequency (right panel). The color on the right plots corresponds to an evaluation time that matches the color on the left plots.}\label{GWplots}
\end{figure}

\noindent
The (normalized) GW energy density per logarithmic frequency interval can be expressed as
\begin{equation}
\Omega_\text{GW} (p,\tau) \, = \, \dfrac{1}{\rho_c(\tau)} \dfrac{\diff  \rho_\text{GW}(p,\tau)}{\diff \log p}\,,
\label{eq:normalizedGWspectrum}
\end{equation}
where $\rho_\text{GW}$ is the total GW energy density and $\rho_c$ is the (time-dependent) critical energy density defined via $\rho_c \equiv 3 H^2M_P^2 $. From the lattice we can infer the full spectral and time dependence of the normalized GW energy-density in Eq.~(\ref{eq:normalizedGWspectrum}). The total GW energy density relative to the critical density at a given time can be computed in a straightforward way
\begin{equation}
   \dfrac{\rho_\text{GW}(\tau)}{\rho_c(\tau)} \, = \, \int \diff \log p \, \Omega_\text{GW} (p,\tau) .
\end{equation}
 The right-hand-side is also estimated at a given time by \CL~as an average over the lattice volume $V$ (see Ref.~\cite{Figueroa:2021yhd} for details) and reads
\begin{equation}
\dfrac{\rho_\text{GW}(\tau)}{{\rho_c(\tau)}} \, = \, \dfrac{1}{\rho_c(\tau)}\dfrac{M_P^2}{4 a^4}\dfrac{1}{V} \int \diff^3 \bp \, \bar{h}_{ij}^\prime (\tau,\bp) \, \bar{h}_{ij}^{\prime *}(\tau,\bp) \,,
\end{equation}
with $\bar{h}_{ij} \equiv a h_{ij}$. This quantity is represented as a function of the scale factor in Fig.~\ref{GWplots} on the left panels for $k=4,6,8,10$. One can notice a sharp increase of the relative gravitational wave energy density for $a\simeq a_\text{frag}$ when non-linearities become significant. The spectral dependence of the GW energy density is represented in Fig.~\ref{GWplots} on the right panels in terms of the present day frequency. The colors of the various curves on the right panels code the evaluation time which can be inferred from the left panels. From the right panels of Fig.~\ref{GWplots} one can see scales prone to parametric resonances progressively redshift towards the infrared with time for $k\neq 4$,\,\footnote{As mentioned, the time-independence of the resonance structure for $k=4$ is a consequence of the conformal invariance of the potential being quartic about the minimum. See Ref.~\cite{Garcia:2023eol} for details.} and being imprinted in the primordial tensor spectrum. While the resonance structure for $k=4$ can be observed in the final spectrum, which can be estimated from the Boltzmann approximation~\cite{Garcia:2023eol}, the time dependence of the resonance bands for $k=6,8,10$, and the subsequent non-linearities in the scalar perturbations, would eventually erase any remnant of such resonances by generating a relatively broad spectrum. This effect was already noticed at the level of the inflaton-fluctuation occupation number in Ref~\cite{Garcia:2023dyf}. \par \medskip

\noindent
The relative energy density after fragmentation, deep into the inflaton-quanta dominated era at $a=a_\text{RD}$,\,\footnote{In practice we take $a_\text{RD}$ to be the scale factor evaluated at the final simulation time.} can be deduced from the asymptotic values reached on the left panels of Fig.~\ref{GWplots} and correspond to 
\begin{equation}
    \left(\dfrac{\rho_\text{GW,RD}}{\rho_{c,\text{RD}}} \right) \, \simeq \, \begin{cases}
7.3 \times 10^{-6}  \, \qquad [k=4],\\
2.5 \times 10^{-7} \, \qquad [k=6],\\
7.6 \times 10^{-8} \, \qquad [k=8],\\
1.6 \times 10^{-8} \,  \qquad [k=10].
\end{cases}
\end{equation}
which is the largest for $k=4$ and decreases as $k$ increases. This quantity can be related to the present GW energy density 
\begin{equation}
\rho_\text{GW,0} \, = \, \rho_\text{GW,RD} \left( \dfrac{a_\text{RD}}{a_0}\right)^4 \, = \,  \rho_\text{GW,RD}  \left( \dfrac{\rho_{\text{rad},0}}{\rho_{c,\text{RD}}} \right) \left( \dfrac{g_\text{reh}}{g_0} \right) \left( \dfrac{g_{s,0}}{g_{s,\text{reh}}} \right)^{4/3}\,.
\end{equation}
The present day normalized GW energy density can thus be expressed as
\begin{equation}
\Omega_{\text{GW},0} (f) \, = \, \dfrac{\rho_{\text{rad},0}}{\rho_{c,0}}  \left( \dfrac{g_\text{reh}}{g_0} \right) \left( \dfrac{g_{s,0}}{g_{s,\text{reh}}} \right)^{4/3}   \, \Omega_\text{GW,RD}(f) \, \simeq \,   \, 2 \times 10^{-5} \, \Omega_\text{GW,RD}(f) \,,
\end{equation}
where we took $g_\text{reh}=g_{s,\text{reh}}=106.75$, $g_{s,0}=3.909$ and $g_0=3.363$. The total energy density at the present day is
\begin{equation}
    \Omega_{\text{GW},0} h^2 \, = \, h^2 \int \diff \log f \,\Omega_{\text{GW},0} (f) \, \simeq \, 10^{-5}  \left(\dfrac{\rho_\text{GW,RD}}{\rho_{c,\text{RD}}} \right) \,.
\end{equation}

The energy density corresponding to the gravitational wave signal can potentially increase the effective number of relativistic species $\Delta N_\text{eff}\equiv N_\text{eff}-N_\text{eff}^\text{SM}$ with respect to the Standard-Model expected value $N_\text{eff}^\text{SM}=3.046$. This contribution can be estimated as
\begin{equation}
\Delta N_\text{eff} \, = \, \dfrac{8}{7} \left(\dfrac{11}{4} \right)^{4/3}  \left(\dfrac{\rho_\text{GW,0}}{\rho_{\text{rad},0}} \right) = \, \dfrac{8}{7} \left(\dfrac{11}{4} \right)^{4/3}  \left( \dfrac{g_\text{reh}}{g_0} \right) \left( \dfrac{g_{s,0}}{g_{s,\text{reh}}} \right)^{4/3} \left(\dfrac{\rho_\text{GW,RD}}{\rho_{c,\text{RD}}} \right) \, \simeq \, 1.7  \left(\dfrac{\rho_\text{GW,RD}}{\rho_{c,\text{RD}}} \right)\,,
\end{equation}
yielding $\Delta N_\text{eff}$ at most of order $\mathcal{O}(10^{-5})$ for $k=4$ and smaller by one or two orders of magnitude for larger values of $k$. Such small values of $\Delta N_\text{eff}$ are far below current and upcoming sensitivity such as COrE~\cite{thecorecollaboration2011core} or Euclid~\cite{laureijs2011euclid} that are expected to reach $\Delta N_\text{eff} < 0.013$ at the $2\sigma$ level.

\subsection{Tensor modes excitation from inflation} \label{sec:gws2}

As discussed above, during slow-roll inflation the source term in the equation of motion for the tensor modes, i.e. the right-hand side of Eq.~(\ref{eq:tensorEOM}), can be neglected and the equation of motion reduces to its homogeneous form
\begin{equation}
h^{\prime \prime}_{ij}(\bp,\tau)+ 2 \mathcal{H} h^{\prime }_{ij}(\bp,\tau)+p^2 h_{ij}(\bp,\tau) \, = \, 0\,,
\end{equation}	
The rescaled metric perturbation $\bar{h}_{ij}\equiv a h_{ij}$ can be decomposed as a sum over polarization states  
\begin{equation}
    \bar h_{ij}(\bp,
    \tau) \, = \, \sum_{\lambda=+,\times} \bar h_\lambda(\tau,\bp)\, \epsilon_{ij}^\lambda(\bp)  \, .
\end{equation}
where $\epsilon_{ij}^\lambda(\bp) $ are the normalized spin-2 polarization tensors satisfying $\sum_{ij}\epsilon_{ij}^\lambda (\epsilon_{ij}^{\lambda^\prime}) =2 \delta^{\lambda \lambda^\prime}$. Each polarization component of the rescaled tensor mode satisfies then the following oscillator equation
\begin{equation}
   \bar h_\lambda^{\prime \prime} + \omega_p^2 \bar h_\lambda \, = \, 0 \,,
\end{equation}
with the corresponding frequency given by $\omega_p^2 = p^2 - a''/a $. By assuming Bunch-Davies initial conditions for the mode functions deep inside the horizon ($|\tau_0\omega_p|\gg 1$),
\beq
\bar h_\lambda(\tau_0) \;=\; \frac{1}{\sqrt{2\omega_p}} \,, \quad \bar h_\lambda'(\tau_0) \;=\; -\frac{i\omega_p}{\sqrt{2\omega_p}}\,,
\eeq
one can solve the mode-function equation and show that tensor perturbations become frozen after horizon crossing $p\simeq \mathcal{H}\equiv a H$ yielding $h_\lambda \sim p^{-3/2}$ at the end of inflation and a corresponding power spectrum 
\begin{equation}
    P_h \, \equiv \, \dfrac{p^3}{2 \pi^2} \sum_\lambda |h_\lambda|^2 \, \simeq \,  \left. \dfrac{2H^2}{\pi^2 M_P^4} \right|_{p= aH} \,,
\end{equation}
evaluated at horizon crossing. The result, as it is well known, is an approximately scale-invariant power spectrum which remains frozen for superhorizon scales after the end of inflation regardless of the EOS of the universe.\par \medskip

\noindent
\textbf{Short wavelengths.}
 Once tensor perturbations re-enter the horizon, the mode function starts oscillating with a decaying envelope $h_\lambda \sim p^{-3/2} (a_p/a)$ where $a_p$ is the scale factor at horizon re-entry $p\equiv\mathcal{H}(a_p)$. This induces a spectral distortion of the frozen spectrum~\cite{Giovannini:1998bp,Figueroa:2018twl,Barman:2022qgt} through the time dependence of the conformal Hubble rate $\mathcal{H} \sim a^{-2(k-1)/(k+2)}$ implying $a_p \sim p^{-(k+2)(2k-2)}$ and the corresponding energy-density spectrum~\cite{Saikawa:2018rcs}
\begin{equation}
    \Omega_{\text{GW}} h^2 \, \simeq \, \dfrac{1}{24} \Omega_{\text{rad}} h^2 \left( \dfrac{g_\text{reh}}{g_0} \right) \left( \dfrac{g_{s,0}}{g_{s,\text{reh}}} \right)^{4/3} \left( \dfrac{p}{\mathcal{H}} \right)^2   \mathcal{T}_p(a) P_h
\end{equation}
where $\mathcal{T}_p(a) = (a_p/a)^2 \sim p^{-(k+2)(k-1)} $ is the transfer function accounting for the time evolution of the tensor spectrum from its frozen value. This implies an energy-density spectrum scaling as $\Omega_{\text{GW}} h^2 \sim p^{(k-4)/(k-1)} $.  For scales re-entering during radiation domination ($p<p_\text{frag}$) or during inflaton-quanta domination with $k=4$, this yields a flat energy-density spectrum
\begin{equation}
    \Omega_{\text{GW},0} h^2 \, \simeq \, 5.6 \times 10^{-18} \left( \dfrac{V^{1/4}(\phi_*)}{8.1 \times 10^{15}~\text{GeV}} \right)^4 \,\qquad[p<p_\text{frag}],
\end{equation}
where the potential is evaluated at the field value $\phi_*$ reached at the time the CMB pivot scale crosses the horizon during inflation which can be inferred from Eq.~(\ref{eq:phistar}) for a given $k$.\footnote{Given our normalization choice for $\lambda$, the value $V(\phi_*)$ is the same for all $k$.} For scales $p>p_\text{frag}$ entering during an inflaton-quanta dominated universe, the spectrum becomes blue tilted from the momentum dependence of the transfer function up to the largest scales that experienced a super-horizon phase during inflation $k_\text{end}$ which are represented in Fig.~\ref{scales}. Switching to present-day frequency via Eq.~(\ref{eq:translatektof}), the total spectrum can therefore be expressed as
\begin{equation}
    \Omega_{\text{GW},0} h^2 \, \simeq \, 5.6 \times 10^{-18} \left( \Theta \Big[ f_\text{frag}-f \Big] +\Theta \Big[ f-f_\text{frag} \Big] \left( \dfrac{f}{f_\text{frag}} \right)^{(k-4)/(k-1)} \right)  \Theta \Big[ f_\text{end} -f \Big] \,,
    \label{eq:Omegah2GWTM}
\end{equation}
with $f_\text{frag}$ and $f_\text{end}$ given by Eq.~(\ref{eq:ffragandfend}) for each value of $k$. \par \medskip

\noindent
\textbf{Long wavelengths.} On much larger physical scales, such as the CMB fiducial scale $k_*=0.05~\text{Mpc}^{-1}$, primordial tensor perturbations are customarily parametrized in terms of the tensor-to-scalar ratio which can be expressed as~\cite{Garcia:2020wiy}
\begin{equation}
    r \, \simeq \, \dfrac{12}{N_*^2} \, \simeq \, 3.9 \times 10^{-3} \, \left( \dfrac{55}{N_*} \right)^2 \,,
    \label{eq:r}
\end{equation}
Such tensor perturbations could leave traces in $B-$polarization modes of the CMB and could potentially be detected by a future experiment. The predicted values of $r$ from Eq.~(\ref{eq:r}) appears to be close to the sensitivity reach for the upcoming Simons Observatory (SO) $r<6 \times 10^{-3}$~\cite{SimonsObservatory:2018koc}. Moreover, future missions such as LiteBIRD and CMB-S4 should allow to disprove or confirm this model by reaching sensitivities as large as $r<2\times 10^{-3}$~\cite{LiteBIRD:2022cnt}  and $r< 10^{-3}$~\cite{SO-CMBS4} respectively. \par \medskip

\subsection{Additional contributions to gravitational wave production}\label{sec:gws3}

There have been substantial developments in the computation of gravitational wave production from the reheating dynamics where various contributions have recently been estimated: 
\begin{itemize}
\item GW from graviton bremsstrahlung during reheating, from inflaton decay or inflaton annihilation~\cite{Nakayama:2018ptw,Huang:2019lgd,Barman:2023ymn,Bernal:2023wus,Barman:2023rpg}. These contributions can be important for large reheating temperatures but suppressed at low reheating temperatures. We expect them to be subdominant in our case.
    \item Prompt gravitational waves sourced by the oscillating inflaton condensate~\cite{Choi:2024ilx}. This contribution typically results in a spectrum $\Omega_{\text{GW},0} h^2 \simeq 10^{-15} (f/f_\text{end})^{(4k-7)/(k-4)}$ peaked close to the maximal frequency $f_\text{end}$ and decreasing sharply for $f<f_\text{end}$. 
\end{itemize}
These contributions are thus expected to be subdominant compared to the signal estimated in this work.

\subsection{Results and discussion}
\label{sec:summary}

\begin{figure}[t!]
\centering{
\includegraphics[width=0.9 \linewidth]{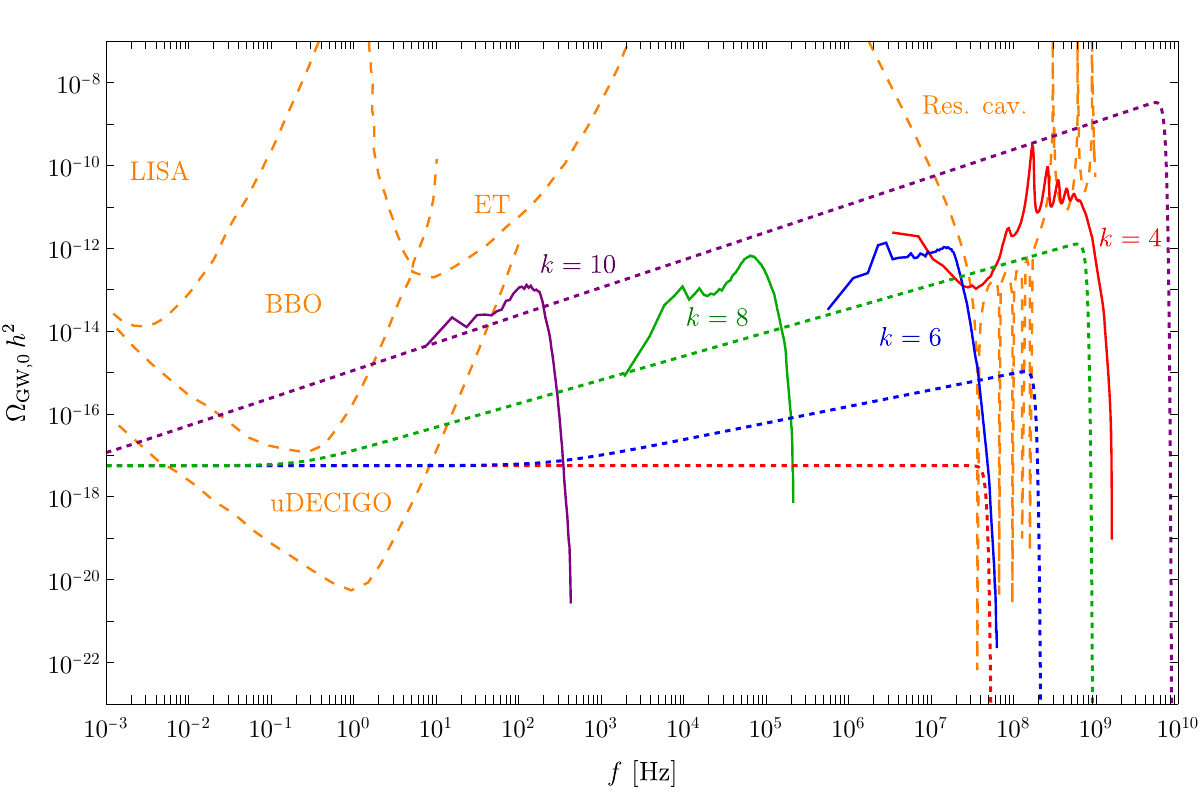}
}
\caption{Total stochastic gravitational wave background as a function of the present-day frequency. Purple, green, blue and red curves correspond to $k=10,8,6,4$ respectively. The solid lines represent the GW signal from non-linearities during the fragmentation process and dotted lines GW resulting from tensor mode super-horizon excitation during inflation. Orange-dashed line are sensitivity prospects for future experimental facilities, see Sec.~\ref{sec:summary} for details.}\label{GWfinalplot}
\end{figure}

The gravitational wave spectra resulting from tensor modes excited during inflation, i.e. Eq.~(\ref{eq:Omegah2GWTM}), are represented for each value of $k=4,6,8,10$ in Fig.~\ref{GWfinalplot} as dotted lines. The gravitational wave signals sourced by non-linearities from the fragmentation process, i.e. the red lines in Fig.~\ref{GWplots} right panels, are in turn shown as solid lines. We also represented sensitivity prospects for future gravitational wave observatories, recently reviewed in~\cite{Aggarwal:2020olq}: the Big Bang Observer (BBO)~\cite{Crowder:2005nr,Corbin:2005ny}, the (ultimate) DECi-hertz Interferometer Gravitational wave Observatory (uDECIGO)~\cite{Seto:2001qf,Kudoh:2005as,Yagi:2011wg}, the Laser Interferometer Space Antenna (LISA)~\cite{2017arXiv170200786A}, the Einstein Telescope (ET)~\cite{Sathyaprakash:2012jk,Hild:2010id,Maggiore:2019uih} in addition to estimate from resonant cavities (Res.~cav.) of Ref.~\cite{Herman:2022fau}. This figure illustrates the complementary between $\mathcal{O}$(Hz) detectors such as BBO or DECIGO and $\mathcal{O}$(GHz) cavities in probing gravitational wave production in our setup. \par \medskip

\noindent
\textbf{Hz detectors.} BBO should almost reach the sensitivity required to detect the scale-independent part of the spectrum at low frequency for $k=4,6,8$ but should be able to detect the $\Omega_{\text{GW},0} h^2 \simeq 10^{-15}\,(f/\text{Hz})^{2/3}$ signal for $k=10$. The sensitivity of uDECIGO should be large enough to probe the scale invariant part of the spectrum for all values of $k$ considered in addition to the $\Omega_{\text{GW},0} h^2 \simeq 10^{-17}\,(f/\text{Hz})^{4/7}$ signal for $k=8$ and the GW generated from fragmentation around $f\sim 10~\text{Hz}$ for $k=10$. \par \medskip

\noindent
\textbf{GHz detectors.}  Electromagnetic cavities have recently garnered interest as a possibility to detect high-frequency gravitational waves. Future experimental setups could pave the way for a new frontier in GW detection~\cite{Berlin:2021txa}. Potential high frequency GW detectors with a sensitivity as high  as proposed in Ref.~\cite{Herman:2022fau} should be able to resolve the peaked structure of the spectrum generated for $k=4$, pointing undoubtedly towards an inflaton potential quartic about the minimum. Moreover, such experiment could detect tensor modes excited during inflation and re-entering the Hubble radius during the coherently-oscillating inflaton era for both $k=10$ at $\Omega_{\text{GW},0} h^2\sim 10^{-10}$ and $k=8$ at $\Omega_{\text{GW},0} h^2\sim 10^{-13}-10^{-12}$. Since both signals are also in the reach of uDECIGO at much lower frequencies, the combination of results from both  experimental facilities could be used to identify or disprove the underlying inflation scenario. Nevertheless, fragmentation signals for $k=6$ and $k=8$ appears not to be in the reach of future experimental facilities.

\section{Summary and conclusion}
\label{sec:conclusions}

In this work we have investigated a consistent picture of the universe describing its early states, starting from inflation and arriving to a phase dominated by a bath of ultra-relativistic particles. We considered for definiteness the T-model of inflation, characterized by an exponent $k$ that we assigned to $k=4,6,8,10$ as specific realizations. While cosmic inflation, with an equation of state $w\simeq -1$, is sensitive to the asymptotically flat part of the inflaton potential at large field values, the exponent $k$ controls the expansion history close to the minimum, where $V(\phi)\sim \phi^k$. This post-inflation history is additionally determined by the inflaton coupling to states that eventually make up the primordial radiation bath. In this work we explored the possibility that this coupling is sufficiently small that reheating would only occur after a significant amount of time posterior to inflation, i.e. late reheating. Immediately after inflation, coherent oscillations of the inflaton field about the minimum of its potential generate a phase of expansion with an oscillation-averaged equation-of-state $w\simeq (k-2)/(k+2)$. Subsequently, excited by the oscillations, parametric resonant effects induce the growth of inhomogeneities in the form of copiously produced inflaton quanta eventually reaching the non-linear regime. Significant mode-mode couplings trigger the fragmentation of the inflaton condensate, marking the transition between a coherently-oscillation phase to an era dominated by a collection of relativistic inflaton quanta, corresponding to an equation of state $w\simeq 1/3$. Under the assumption of weak couplings, the transition to a universe dominated by a Standard Model thermal bath in equilibrium (i.e. reheating) is therefore achieved only once fragmentation is completed. This transition leaves the equation of state unaffected $\simeq 1/3$, therefore the precise value of the reheating temperature does not affect the cosmological history, and the subsequent phenomenological consequences explored in this work. \par \medskip

\noindent
In this paper, we kept track of the expansion history by solving the equation of motion for the inflaton field during inflation and by simulating the post-inflation history with~\CL. We estimated the time dependence of the equation-of-state during the fragmentation process of the inflaton. While fragmentation is relatively fast for $k=4$, occurring at $a_\text{frag}/a_\text{end}\simeq 180$, it takes longer to be achieved as the value of $k$ increases, yielding 
$a_\text{frag}/a_\text{end} \simeq 4.5 \times 10^4,6 \times 10^6,7\times 10^8$  for $k=6,8,10$ and a longer period of coherent-oscillations for large values of $k$. \par \medskip

\noindent
We estimated the gravitational wave signal generated during the cosmological history in our setup and identified two main contributions. The first contribution consists of tensor modes crossing the horizon and being excited during inflation. Such modes could re-enter during a phase of the universe with an equation of state $w=1/3$ such as inflaton-quanta or radiation dominated phase yielding a scale invariant signal $\Omega_{\text{GW},0} h^2\simeq 5.6 \times 10^{-18}$ for $k=4,6,8,10$. Modes re-entering during the coherent inflaton-oscillation phase would benefit from a stiff equation of state $w=(k-2)/(k+2)$ inducing a blue-tilted signal $\Omega_{\text{GW},0} h^2 \sim f^{(k-4)/(k-1)} $. While the transition between the scale-invariant and the blue-tilted spectra is determined by the size of the Hubble radius at fragmentation, the ultra-violet cutoff corresponds to the Hubble radius at the end of inflation. The second contribution to the GW signal comes from the large inhomogeneities generated during the fragmentation process sourcing tensor perturbations at second order in cosmological perturbation theory. This results in a high frequency signal for $k=4$ with a peaked structure at around $f\sim 10^8~\text{Hz}$ as remnants of the parametric resonance structure. For larger values of $k=6,8,10$, non-linearities induce broad spectra at frequencies respectively of order $f\sim  10^7,  10^5,10^2~\text{Hz}$. \par \medskip

\noindent
We compared our results to prospects for experimental facilities sensitive to gravitational waves. We identified three possibilities of probing the gravitational wave signals in the future. First, the scale-invariant part of the spectrum on very large physical scales could be detected in the form of primordial tensor modes by LiteBIRD and CMB-S4 experiments for all $k=4,6,8,10$ considered. Second,  uDECIGO should reach a sensitivity sufficient to detect the scale-invariant spectrum for frequencies $f \sim \text{Hz} $ for $k=4,6$  and the blue-tilded component of the spectrum for $k=8,10$ as well as the fragmentation-induced signal for $k=10$. At last, resonant electromagnetic cavities sensitive to frequencies $f\sim10^8~\text{Hz}$ should be able to detect the blue-tilted spectra for $k=8,10$ and the fragmentation signal for $k=4$. \par \medskip

\noindent
Our work shows the ability for future experimental facilities to detect gravitational wave signals emitted in the early universe as well as the complementarity between detectors sensitive to different scales corresponding to different epoch of the universe. Gravitational wave astronomy will be an essential probe of inflation scenario and determinant to discriminate various interpretations in the decades to come. While strong efforts are currently employed to develop high-frequency GW detectors~\cite{Berlin:2021txa}, we advocate the necessity of building such ultra-sensitive GW observatories to shed light on the dynamics of inflation and reheating.

\begin{acknowledgments}

 MG is supported by the DGAPA-PAPIIT grant IA103123 at UNAM, the CONAHCYT “Ciencia de Frontera” grant CF-2023-I-17, and the PIIF-2023 grant from Instituto de F\'isica, UNAM. MP acknowledges support by the Deutsche Forschungsgemeinschaft (DFG, German Research Foundation) under the DFG Emmy Noether Grant No. PI 1933/1-1 and Germany's Excellence Strategy – EXC 2121 “Quantum Universe” – 390833306. 
\end{acknowledgments}

\addcontentsline{toc}{section}{References}
\bibliographystyle{utphys}
\bibliography{references}

\end{document}